\documentclass[twocolumn]{aastex61}
\pdfoutput=1 
\usepackage{amsmath,amstext}
\usepackage[T1]{fontenc}
\usepackage{apjfonts} 
\usepackage[figure,figure*]{hypcap}
\newcommand{\ow}{OW\,J0741}
\newcommand{\kms}{\ensuremath{{\rm km}\,{\rm s}^{-1}}}
\newcommand{\degree}{$^{\circ}$ }
\newcommand{\msol}{M$_\odot$}
\newcommand{\teff}{$T_{\rm eff}\text{ }$}
\newcommand{\logg}{$\log{g}\text{ }$}
\newcommand{\porb}{P$_{\rm orb}$}

\shorttitle{The ultracompact sdOB binary \ow}
\shortauthors{T. Kupfer et al.}

\begin{document}

\title{The OmegaWhite survey for short-period variable stars - V. Discovery of an ultracompact hot subdwarf binary with a compact companion in a 44\,minute orbit}

\author{T. Kupfer}
\affiliation{Division of Physics, Mathematics and Astronomy, California Institute of Technology, Pasadena, CA 91125, USA}
\author{G. Ramsay}
\affiliation{Armagh Observatory and Planetarium, College Hill, Armagh BT61 9DG, UK}
\author{J. van Roestel}
\affiliation{Department of Astrophysics/IMAPP, Radboud University, PO Box 9010, NL-6500 GL Nijmegen, the Netherlands}
\author{J. Brooks}
\affiliation{Department of Physics, University of California, Santa Barbara, CA 93106, USA}
\author{S. A. MacFarlane}
\affiliation{Department of Astrophysics/IMAPP, Radboud University, PO Box 9010, NL-6500 GL Nijmegen, the Netherlands}
\affiliation{Department of Astronomy, University of Cape Town, Private Bag X3, Rondebosch 7700, South Africa}
\author{R. Toma}
\affiliation{Armagh Observatory and Planetarium, College Hill, Armagh BT61 9DG, UK}
\author{P. J. Groot}
\affiliation{Department of Astrophysics/IMAPP, Radboud University, PO Box 9010, NL-6500 GL Nijmegen, the Netherlands}
\affiliation{Kavli Institute for Theoretical Physics, University of California, Santa Barbara, CA 93106, USA}
\affiliation{Department of Astronomy, University of Cape Town, Private Bag X3, Rondebosch 7700, South Africa}
\author{P. A. Woudt}
\affiliation{Department of Astronomy, University of Cape Town, Private Bag X3, Rondebosch 7700, South Africa}
\author{L. Bildsten}
\affiliation{Department of Physics, University of California, Santa Barbara, CA 93106, USA}
\affiliation{Kavli Institute for Theoretical Physics, University of California, Santa Barbara, CA 93106, USA}
\author{T. R. Marsh}
\affiliation{Department of Physics, University of Warwick, Coventry CV4 7AL, UK}
\author{M. J. Green}
\affiliation{Department of Physics, University of Warwick, Coventry CV4 7AL, UK}
\author{E. Breedt}
\affiliation{Department of Physics, University of Warwick, Coventry CV4 7AL, UK}
\author{D. Kilkenny}
\affiliation{Department of Physics \& Astronomy, University of the Western Cape, Private Bag X17, Bellville 7535, South Africa}
\author{J. Freudenthal}
\affiliation{Institut f\"ur Astrophysik, Georg-August-Universit\"at G\"ottingen, Friedrich-Hund-Platz 1, 37077 G\"ottingen, Germany}
\author{S. Geier}
\affiliation{Institute for Astronomy and Astrophysics, Kepler Center for Astro and Particle Physics, Eberhard Karls University, Sand 1,
72076 T\"ubingen, Germany}
\author{U. Heber}
\affiliation{Dr. Remeis-Sternwarte \& ECAP, Astronomical Institute, University of Erlangen-Nuremberg, Germany}
\author{S. Bagnulo}
\affiliation{Armagh Observatory and Planetarium, College Hill, Armagh BT61 9DG, UK}
\author{N. Blagorodnova}
\affiliation{Division of Physics, Mathematics and Astronomy, California Institute of Technology, Pasadena, CA 91125, USA}
\author{D. A. H. Buckley}
\affiliation{South African Astronomical Observatory, Cape Town, South Africa}
\author{V. S. Dhillon}
\affiliation{Department of Physics \& Astronomy, University of Sheffield, Sheffield, S3 7RH, UK}
\affiliation{Instituto de Astrof\'isica de Canarias (IAC), E-38200 La Laguna, Tenerife, Spain}
\author{S. R. Kulkarni}
\affiliation{Division of Physics, Mathematics and Astronomy, California Institute of Technology, Pasadena, CA 91125, USA}
\author{R. Lunnan}
\affiliation{Division of Physics, Mathematics and Astronomy, California Institute of Technology, Pasadena, CA 91125, USA}
\affiliation{The Oskar Klein Centre \& Department of Astronomy, Stockholm University, AlbaNova, SE-106 91 Stockholm, Sweden}
\author{T. A. Prince}
\affiliation{Division of Physics, Mathematics and Astronomy, California Institute of Technology, Pasadena, CA 91125, USA}

\begin{abstract}
We report the discovery of the ultracompact hot subdwarf (sdOB) binary OW\,J074106.0-294811.0 with an orbital period of \porb=$44.66279\pm1.16\times10^{-4}$\,min, making it the most compact hot subdwarf binary known. Spectroscopic observations using the VLT, Gemini and Keck telescopes revealed a He-sdOB primary with an intermediate helium abundance, \teff=$39\,400\pm500$\,K and \logg=$5.74\pm0.09$. High signal-to-noise ratio lightcurves show strong ellipsoidal modulation resulting in a derived sdOB mass $M_{\rm sdOB}=0.23\pm0.12$\,\msol\,with a WD companion ($M_{\rm WD}=0.72\pm0.17$\,\msol). The mass ratio was found to be $q = M_{\rm sdOB}/M_{\rm WD}=0.32\pm0.10$. The derived mass for the He-sdOB is inconsistent with the canonical mass for hot sbudwarfs of $\approx0.47$\,\msol.

To put constraints on the structure and evolutionary history of the sdOB star we compared the derived \teff, \logg and sdOB mass to evolutionary tracks of helium stars and helium white dwarfs calculated with Modules for Experiments in Stellar Astrophysics (\texttt{MESA}). We find that the best fitting model is a helium white dwarf with a mass of $0.320$\,\msol, which left the common envelope ${\approx}1.1$\,Myr ago, is consistent with the observations. As a helium white dwarf with a massive white dwarf companion the object will reach contact in 17.6\,Myr at an orbital period of 5\,min. Depending on the spin-orbit synchronization timescale the object will either merge to form an R\,CrB star or end up as a stably accreting AM\,CVn-type system with a helium white dwarf donor.

\end{abstract}


\section{Introduction}

Hot subdwarfs (sdOs/sdBs) are low mass He-stars with very thin hydrogen envelopes. They have effective temperatures similar to O and B stars but are several orders of magnitude less luminous due to their small size \citep{heb86,heb09,heb16}. \citet{max01} and \citet{nap04a} showed that $>50\,\%$ of sdB/sdO stars are in compact binaries with orbital periods \porb$<10$\,days. Formation and orbital shrinkage through a common envelope phase is the only way to form such compact binaries \citep{han02,han03,nel10a}.

\begin{table*}[!t]
\centering
\caption{Summary of the observations of \ow.}
\begin{tabular}{lllccl}
\hline\hline
 Date   & UT  & Tele./Inst. & N$_{\rm exp}$ & Exp. time (s) & Coverage (\AA)/Filter \\
  \hline
    {\bf Photometry} &   &     &        \\
   2016-04-14 & 17:47 - 20:29    & SAAO/SHOC    & 325   &  30  &  clear  \\
   2016-04-15 & 17:13 - 19:22   & SAAO/SHOC    &  257  &  30  &  clear  \\
   2016-04-18 & 17:04 - 19:29   & SAAO/SHOC    &  216  &  40   &  clear  \\
  2016-04-19 &  18:28 - 19:58  & SAAO/SHOC    & 136   &  40  &  clear  \\
  2016-11-03 &  14:47 - 15:39  & Keck/LRIS  &  56  & 20 & \textit{g'} \\
  2016-11-28 &  05:30 - 06:32  & Gemini/GMOS & 95 & 23 & \textit{g'} \\ 
  2016-12-05 &  05:37 - 06:39  & Gemini/GMOS & 95 & 23 & \textit{g'} \\
  2016-12-06 &  05:32 - 06:35  & Gemini/GMOS & 95 & 23 & \textit{g'} \\
  2016-12-07 &  05:32 - 07:00  & NTT/ULTRACAM & 207 & 20/40 & \textit{u'} \\
  2016-12-07 &  05:32 - 07:00  & NTT/ULTRACAM & 412 & 10/20 & \textit{g'} \textit{r'} \\
  2016-12-07 &  21:57 - 22:45  & SALT/Salticam & 1470 & 2 & clear \\
     \noalign{\smallskip}
 {\bf Spectroscopy} &     &           \\
  2016-09-20  & 08:37 - 09:27  & VLT/FORS2       &  9  & 300       &  3500 - 6300  \\ 
   2016-11-03 & 06:38 - 08:07  &  Gemini/GMOS    & 20 & 240       &   3500 - 6700 \\
   2016-11-03 & 13:22 - 14:08 & Keck/LRIS       & 10  & 240       &   3400 - 5600 \\
   2016-11-04 & 05:39 - 06:23 &  Gemini/GMOS      & 10 & 240       &   3500 - 6700  \\
   2016-11-28 & 12:00 - 13:38  & Keck/LRIS       & 20  & 240       &   3400 - 5600 \\
   2016-12-29 & 10:00 - 10:41  & Keck/LRIS       & 9  & 240       &   3400 - 5600 \\
   2017-01-26 & 09:15 - 10:01  & Keck/LRIS       & 10  & 240       &   3400 - 5600 \\
 {\bf Swift} &     &           \\
 2016-10-24 & 04:06 - 06:02 & XRT & 1 & 2944 & 0.2-10keV \\
 2016-10-24 & 04:06 - 06:02 & UVOT & 1 & 2922 & 1600 - 2260 (UVW2)\\
\hline
\end{tabular}
\label{tab:observ}
\end{table*}

It has been shown that hot subdwarfs in compact sdB/sdO + white dwarf (WD) binaries with \porb$\lesssim$2\,hours on the exit of the common envelope phase still burn helium when the sdB/sdO fills its Roche Lobe. Due to the emission of gravitational waves, the binary is predicted to shrink until the hot subdwarf star fills its Roche lobe at an orbital period of 16 to 50\,min and starts mass transfer. He-rich material is then transferred to the WD companion with typical mass transfer rates of $\dot{M}\approx10^{-8}\,$\msol yr$^{-1}$ (e.g. \citealt{sav86,tut89,tut90,ibe91,yun08,pie14,bro15}).

Most of the known compact sdB/sdO binaries reside in systems with orbital periods $\geq0.1$\,days where the sdB/sdO will have turned into a carbon/oxygen WD when both components get into contact \citep{kup15a}. So far only two hot subdwarf binaries with a white dwarf companion and an orbital period \porb$<90$\,min are known \citep{ven12,gei13,kup17}. 

Just recently \citet{kup17} discovered the ultracompact sdB+WD binary, PTF1\,J082340.04+081936.5, with \porb=87\,min. The object is close to the limit to start future accretion while the sdB is still burning helium. If the sdB still burns helium when the system gets into contact, helium-rich material will be accreted onto the WD companion and the most likely outcome is a helium accreting AM\,CVn-type system because the companion is by then a low mass white dwarf ($M_{\rm WD}=0.46^{+0.12}_{-0.09}$\,\msol). 

The most compact sdB binary with a WD companion is CD-30$^{\circ}$11223 (\porb=70.5\,min, \citealt{ven12,gei13}). The sdB will fill its Roche Lobe in $\approx40$\,million years, well within the He-burning lifetime of the sdB star. The WD companion is massive ($M_{\rm WD}\approx0.75$\,\msol) and after accreting 0.1\,\msol, He-burning is predicted to be ignited unstably in the accreted helium layer on the surface of the white dwarf \citep{bro15,bau17}. This could trigger the ignition of carbon in the core which might disrupt the WD even when the mass is significantly below the Chandrasekhar mass, a so-called double detonation supernova type Ia (e.g. \citealt{liv90,liv95,fin10,woo11,wang13,she14}). The hot subdwarf will become unbound and ejected with the orbital velocity which can be up to $\approx1000$\,\kms. The hypervelocity sdO star US708 \citep{hir05} has been proposed as candidate for such a donor remnant \citep{gei13,gei15}. 

\citet{bil07} showed that unstable He-shell burning can detonate the He-shell without disrupting the WD which can be observed as a faint and fast .Ia supernova. The detonation will increase the orbit and the binary system will lose contact. Gravitational wave radiation will decrease the orbit again bring both objects back into contact which can trigger several subsequent weaker flashes \citep{bro15}.

The OmegaWhite survey is a high cadence synoptic survey of the southern Galactic Plane and Galactic Bulge, the main aim of which is to identify ultracompact binaries \citep{mac15,tom16,mac17,mac17a}. OmegaWhite observations are obtained using the VST telescope at the European Southern Observatory's (ESO) Paranal site in Chile. Each field is imaged an average of 38 times over a 2 hour period, with 39 second integrations. Lightcurves are extracted from image differencing and a Lomb-Scargle and AoV analysis is performed on all extracted lightcurves to identify short-period variables. OW J074106.0--294811.0 (\ow\, hereafter) was discovered as a faint ($g=$20.0) blue ($u-g = -1.23$\,mag, $g-r = 0.18$\,mag) source, photometrically variable on a period of $22.6$ min in the OmegaWhite survey \citep{mac15}. It was noted in \citet{tom16} that further observations of OW J0741 showed that it was a  binary with a $44$ min orbital period. Here we report the discovery of \ow\, as an ultracompact sdOB + WD binary with an orbital period of \porb\,= 44.66\,min, making it the most compact hot subdwarf binary known today.

The paper is structured as follows. Section 2 describes the observations used in the analysis. The orbital and binary parameter as well as the atmospheric parameters are described in section 3 and 4. The lightcurve analysis and the system parameters are presented in section 5 and 6. Section 7 discusses the structure and evolutionary history of the system using \texttt{MESA}. A summary and conclusions are given in section 8.

\section{Observations}
\subsection{Optical photometry}
Optical photometric data were obtained using the SAAO 1.9-m Telescope with the Sutherland High Speed Optical Camera (SHOC), Keck/LRIS, Gemini-South/GMOS as well as the 3.5m New Technology Telescope (NTT) with ULTRACAM and the Southern African Large Telescope (SALT) with SALTICAM. 

Photometry from the South African Astronomical Observatory (SAAO) was obtained using the 1.9m telescope and Sutherland High-Speed Optical Camera (SHOC; \citealt{cop13}) which uses an Andor E2V CCD and works in frame-transfer mode (with 1024$\times$1024 active pixels) so that the read-out time is effectively zero. It has been designed to run at up to 20 frames s$^{-1}$ with high-accuracy timing for each frame but, because this object is so faint, typical exposure times were 30-40 seconds (see Table\,\ref{tab:observ}). No filter was used in order to maximize the count rate and the SHOC data were reduced using an in-house SAAO aperture photometry pipeline based on \texttt{IRAF} routines.  On the 1.9m telescope, SHOC has a field-of-view of 2.8$\times$2.8 \arcmin so that several brighter stars were available for differential photometry. We used three photometrically stable field stars for photometric calibration. The observations were conducted under partly cloudy conditions with an average seeing of 1.2\arcsec.

Observations were made of {\ow} on December 7, 2016 using the high speed photometer SALTICAM (see \citealt{odo96}) mounted on SALT at the Sutherland Observatory in South Africa.  Observations were made with no filter and an exposure time of 2 sec and the dataset consists of 1470 images. The SALTICAM frame transfer mode allows for exposures with 200 millisec readout time.

Photometric follow-up observations were obtained with Keck and Gemini-South using LRIS \citep{oke95} and GMOS \citep{hoo04} in imaging mode respectively. Both setups used the \textit{g'} filter and a 20\,sec exposure time. For GMOS we used a 2$\times$2 binning and a 1$\times$1\,arcmin window to reduce the readout time to $\approx$10\,sec. Gemini observed \ow\, at an airmass $\approx1.05$ with an average seeing of 0.7\arcsec. The Keck lightcurve was obtained at an airmass of $\approx1.5$ and an average seeing of 1.1\arcsec. The readout time of the CCD was 42\,sec for each exposure.

High cadence observations were obtained using ULTRACAM \citep{dhi07} on the ESO 3.58m New Technology Telescope (NTT) on December 6, 2016. ULTRACAM is a high-speed photometer that observes in 3 color bands simultaneously and uses frame-transfer CCDs to reduce dead time between exposures to almost zero. Sloan filters \textit{u'}, \textit{g'} and \textit{r'} were used. The exposure time was in the beginning 10\,seconds and later 20~seconds in \textit{g'} and \textit{r'}, and 20~seconds and 40~seconds in \textit{u'} to compensate for the lower throughput of that band. These observations covered two full orbits of the system under clear conditions with an average seeing of 1\arcsec. Each image was bias- and dark-subtracted, and divided through by a twilight flat field.

The photometric data from Keck, Gemini and ULTRACAM were reduced using the ULTRACAM pipeline. The count rate of \ow\, was extracted using differential aperture photometry, using the same comparison star for Keck, Gemini and ULTRACAM to remove atmospheric transparency effects.

An aperture of 1.7~$\times$~the FWHM of the star was used to extract the photometry in the Keck and Gemini data. For the ULTRACAM data, the aperture width was reduced to 1.5~$\times$~the FWHM of the star, in order to avoid contamination from nearby stars due to the lower resolving power of the NTT.

\begin{figure}
\begin{center}
\includegraphics[width=0.49\textwidth]{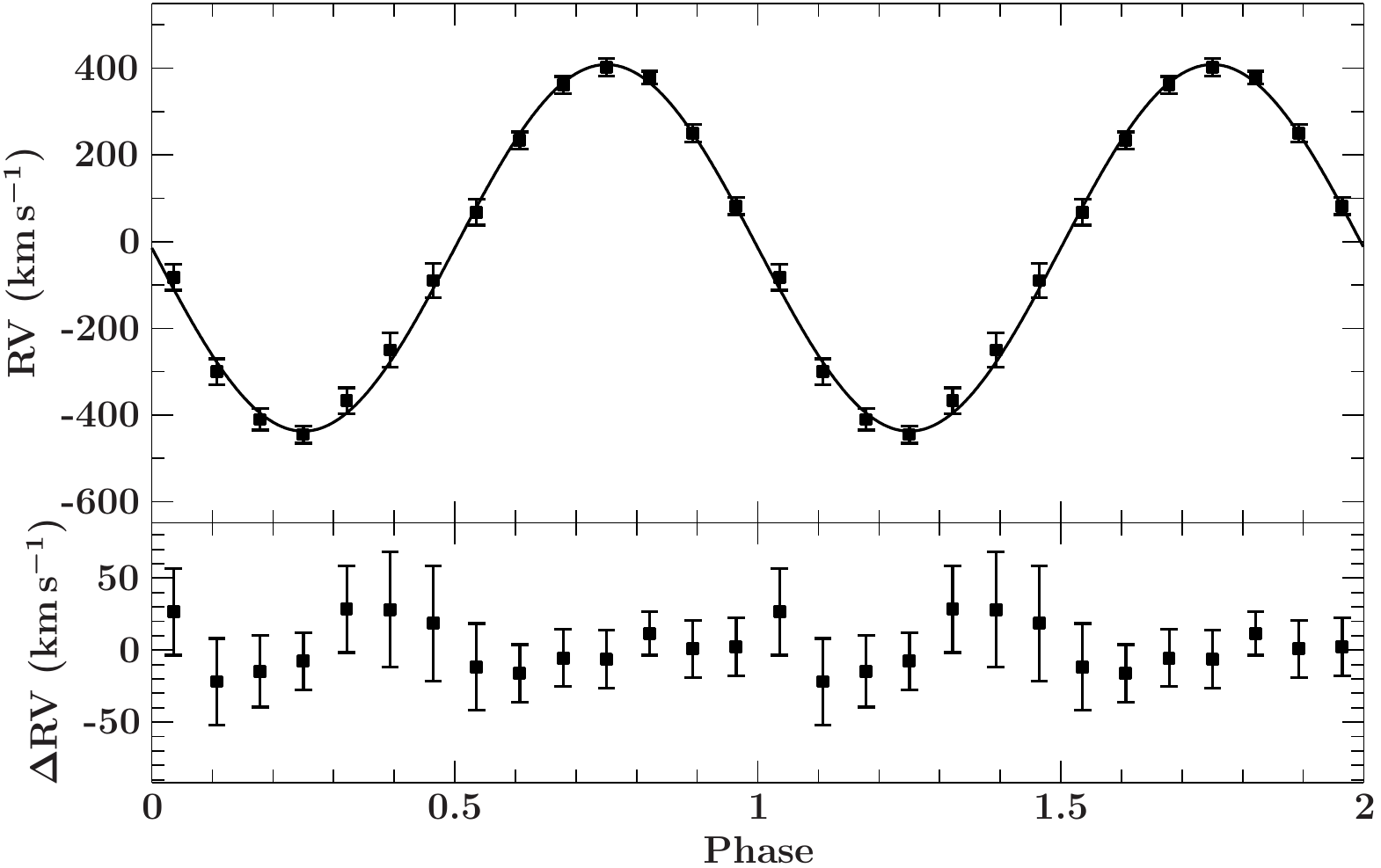}
\end{center}
\caption{Radial velocity plotted against orbital phase. The RV data were phase folded with the orbital period and are plotted twice for better visualization. The residuals are plotted below.}
\label{fig:rv_curve}
\end{figure}

\begin{figure*}
\begin{center}
\includegraphics[width=\textwidth]{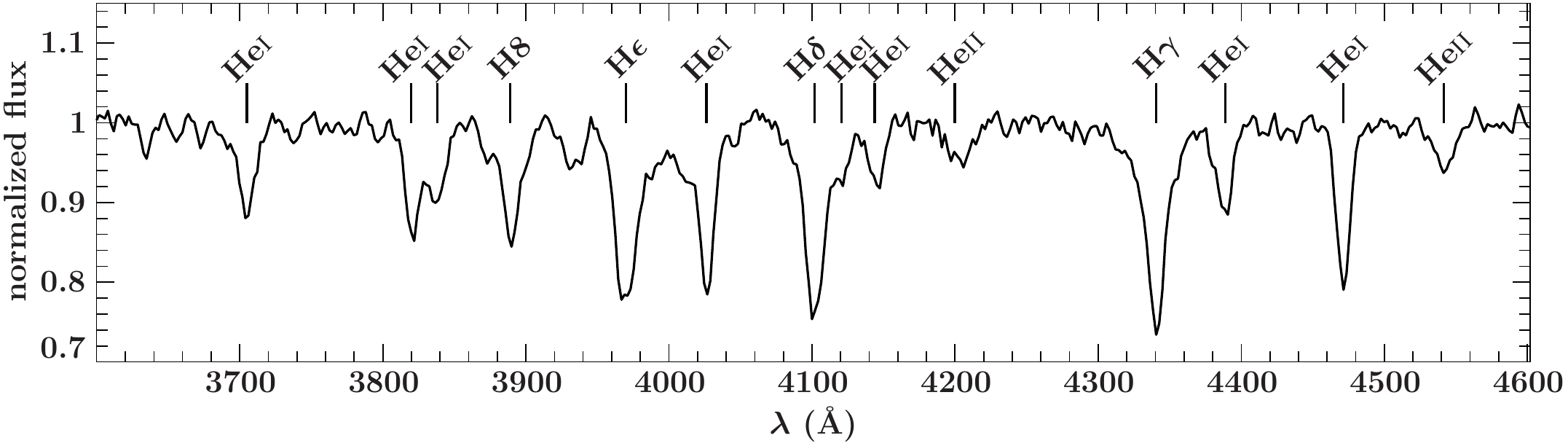}
\includegraphics[width=\textwidth]{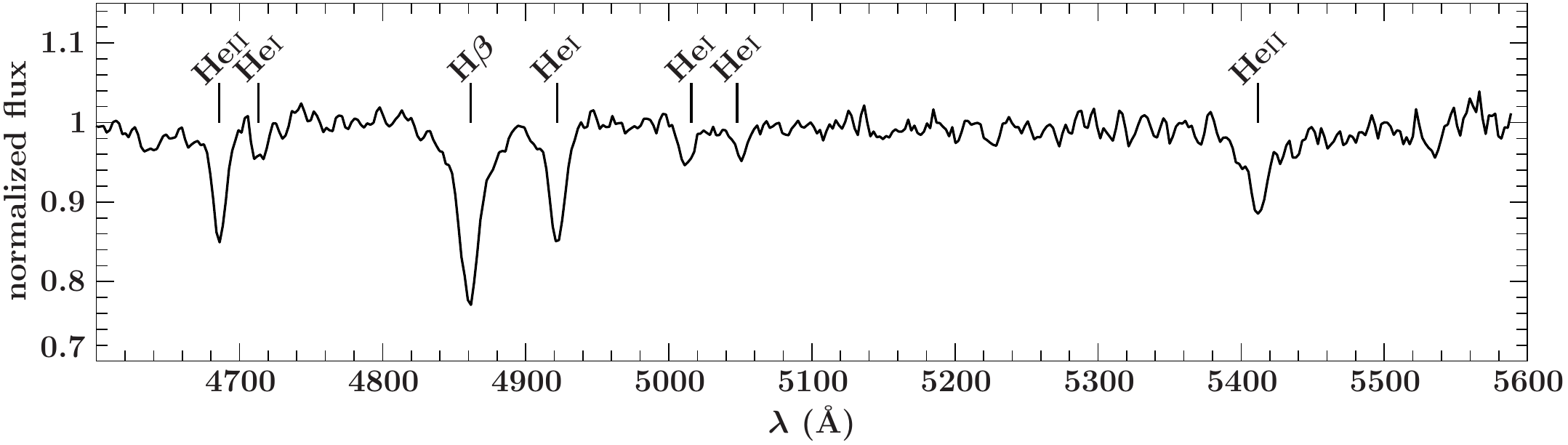}
\end{center}
\caption{Average spectrum of \ow. All prominent lines are marked.}
\label{fig:ow0741_aver}
\end{figure*}

\begin{figure*}
\begin{center}
\includegraphics[width=0.95\textwidth]{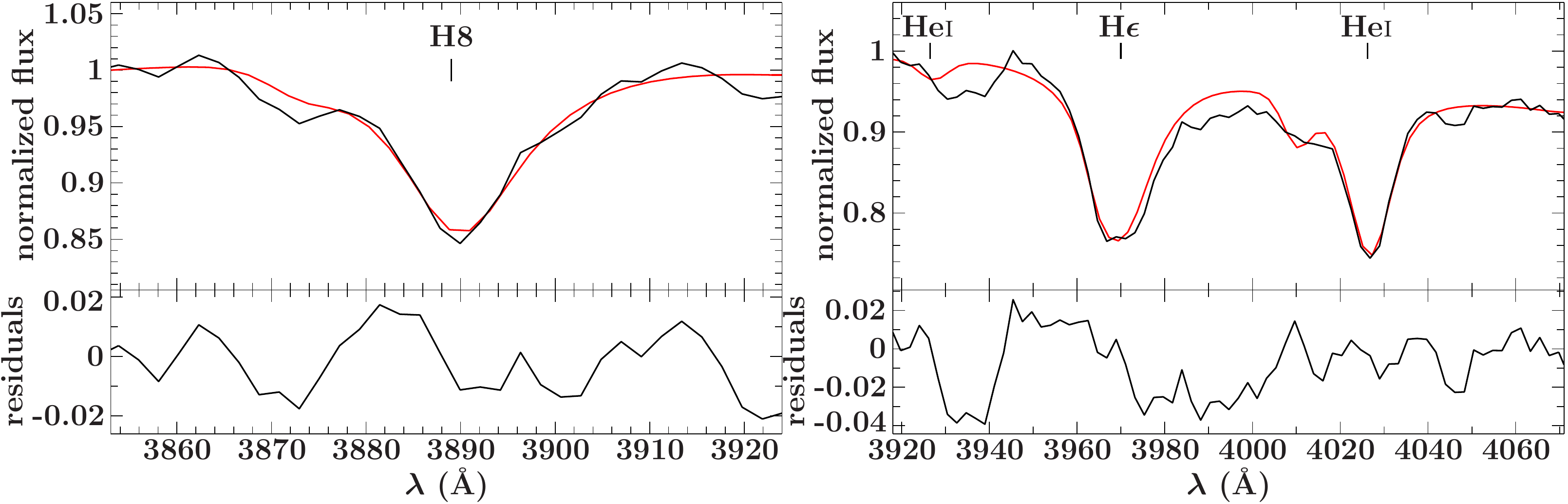}
\includegraphics[width=0.95\textwidth]{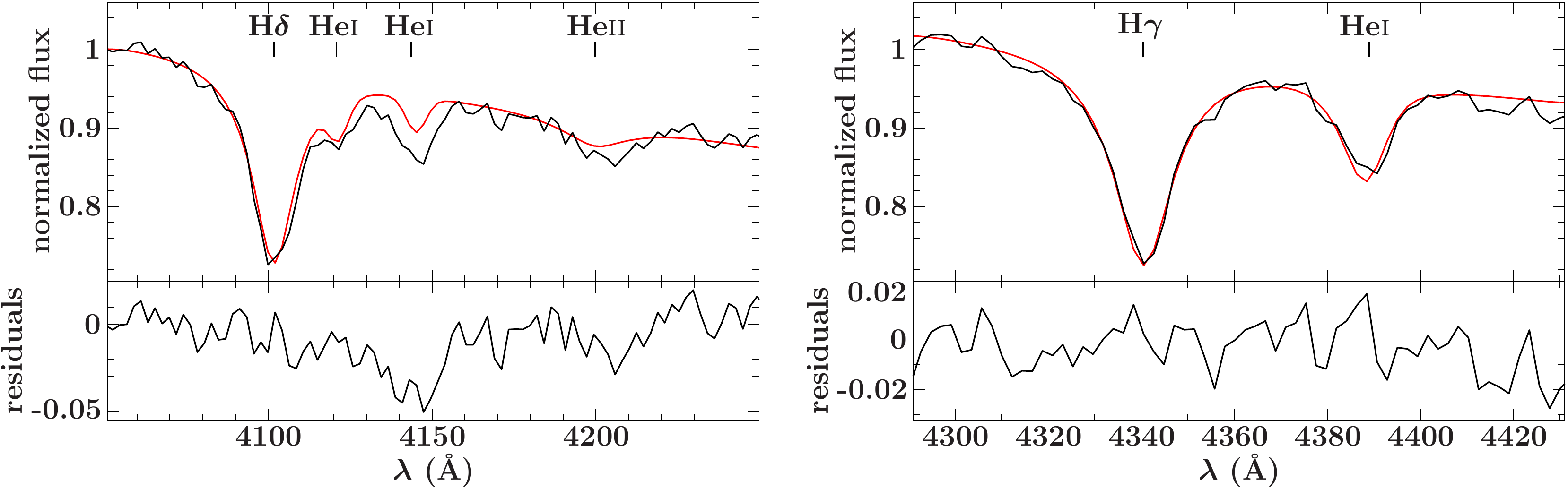}
\includegraphics[width=0.95\textwidth]{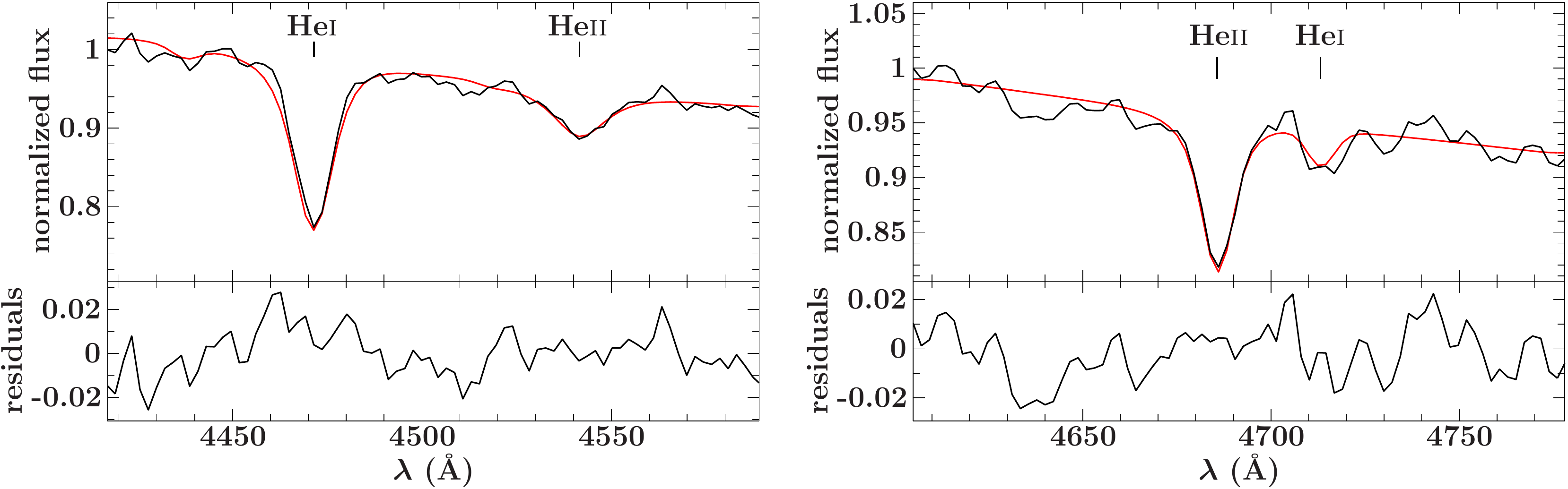}
\includegraphics[width=0.95\textwidth]{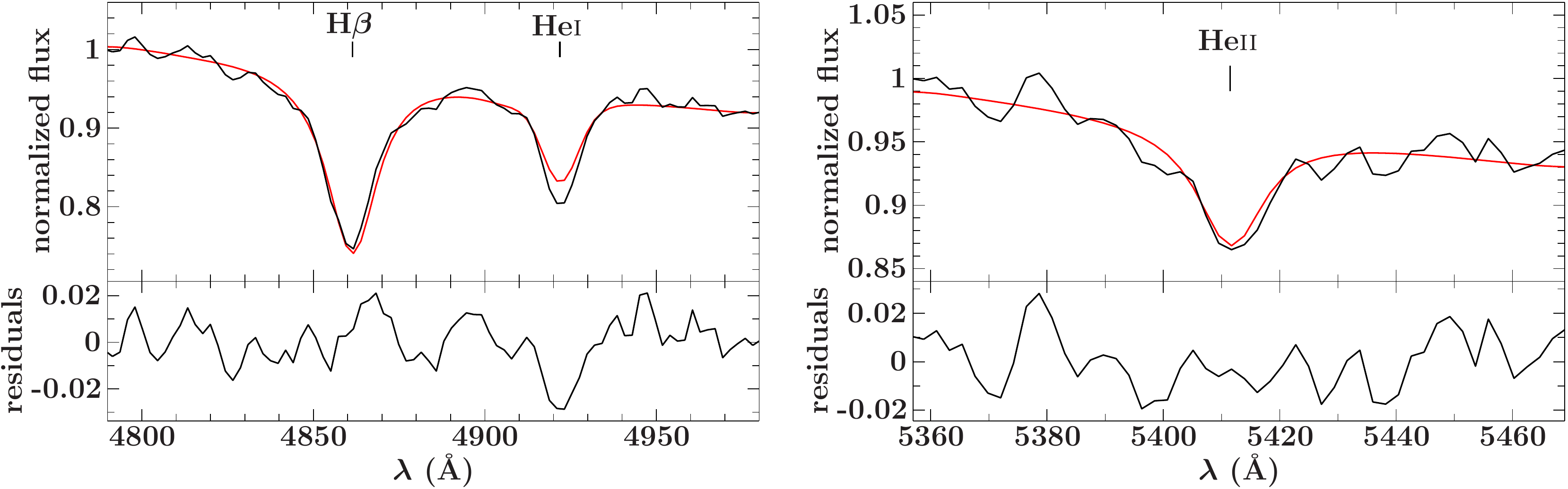}
\end{center}
\caption{Fit of synthetic NLTE models to the hydrogen Balmer, as well as neutral and ionized helium lines of the co-added spectrum.}
\label{fig:ow0741_spec}
\end{figure*}

\subsection{Spectroscopy}

\subsubsection{ESO VLT}

We obtained longslit spectroscopy of \ow\, using the FORS2 instrument (\citealt{app98}) at the ESO VLT UT1.  We used grism
600B with an 1\arcsec\ slit width, for a spectral resolution of $\sim 1000$ (measured on the arc lines), and we obtained 9 spectra on the night 20 September, 2016 between 08:37 and 09:27 UT, each of them with an exposure time of 300\,s. The original pixel size was 1.38\,\AA, but to increase the signal-to-noise (S/N) we smoothed the spectra with a running average 6 pixel wide, effectively reducing the spectral resolution to $\sim 600$. The ambient condition report shows that the night was photometric and that during the observations seeing varied from 1\arcsec\ to 1.25\arcsec. Peak {\it S/N} of individual spectra was $\sim 8$ per \AA.

Data reduction was performed using standard {\sc IRAF} routines. We used spectrophotometric standard stars obtained during nights close to that of the observations of the main target. White dwarf LDS\,749B (= LAWD\,87) was observed on September 18, 2016 at UT 03:26, and white dwarf EGGR\,21 (= CPD$-69$ 177) was observed twice consecutively on September 22, 2016 at UT 08:23 and 08:26 within the context of the standard FORS calibration plan. Calibrated fluxes were taken from \citet{oke90} for LDS\,749B and from \citet{ham92,ham94} for EGGR\,21. The observations of the standard stars were obtained in MOS mode with a slit of 5\arcsec\ width, located in a slightly offset position with respect to the 1\arcsec\ longslit used for the main target, with the consequence that the spectral coverage of the standard star does not perfectly overlap that of target, and transmission function at wavelengths longer than 6000\,\AA\ had to be extrapolated.  Fluxes were corrected for atmospheric extinction using Paranal extinction coefficients from \cite{pat11}. The correction of the instrumental sensitivity obtained with the two different datasets of standard stars led to consistent results, and in the end we adopted only star EGGR\,21 as a calibrator. The use of a 1\arcsec\ slit width while seeing conditions were about 1.25\arcsec\ led to wavelength dependent slit losses up to 30\,\%.

\subsubsection{Gemini South}

Optical spectra were obtained using Gemini-South with the GMOS spectrograph over 2 nights using a low resolution mode ($R\approx1500$). Twenty consecutive spectra were obtained on November 3, 2016 and 10 consecutive spectra were obtained on November 4, 2016. An average bias frame was made out of 5 individual bias frames and a normalized flat-field frame was constructed out of 6 individual lamp flat-fields. CuAr arc exposures were taken before and after 10 consecutive spectra to correct for instrumental flexure. Each exposure was wavelength calibrated by interpolating between the two closest exposures. All spectra were reduced using the \texttt{IRAF} package for GMOS.

\subsubsection{Keck}

Over four nights, we obtained a total of 49 spectra using the Keck I telescope and the LRIS spectrograph \citep{oke95} in a low resolution mode ($R\approx750$). Ten bias frames were obtained to construct an average bias frame and 10 individual lamp flatfields were obtained to construct a normalized flatfield. \mbox{HgNeArCaSn} arc exposures were taken before and after each observing sequence. Each exposure was wavelength calibrated by interpolating between the two closest exposures and skylines were used to correct for instrumental flexure. All spectra were reduced using a custom \texttt{IDL}-based package.

\subsection{X-ray and UV observations}

We obtained Director's Discretionary Time on the {\sl Swift} satellite on October 24, 2016 giving a exposure time of 2944 sec and 2922\,sec in the X-ray Telescope (XRT) and UVOT (UVW2 filter) respectively. The XRT is sensitive over the range 0.2-10\,keV and we examined data taken in `photon counting' mode and used the products derived from the standard XRT pipeline. 

An overview of all our observations is given in Table\,\ref{tab:observ}.

\section{Orbital and binary parameters}\label{sec:orbit}

The photometric data which led to the discovery of \ow\, was a 2.5 hr sequence of $44\times39$\,sec exposures in the \textit{g'}-band as part of the OmegaWhite survey \citep{mac15}. This lightcurve showed a strong peak in its power spectrum at 22.6 min and a corresponding amplitude of 0.22 mag \citep{tom16}. The SAAO 1.9m telescope and SHOC in April, 2016 (see Tab.\,\ref{tab:observ}) show unequal minima on top of the predominantly sinusoidal variability. The dominant modulation in the lightcurve is caused by ellipsoidal deformation of the sdOB and the unequal minima are caused due to gravity darkening of the deformed sdOB. The unequal depth of the minima was confirmed when we obtained 8m class photometric data from Keck and Gemini.

To determine an ephemeris we initially determined the time of the deepest minima in the Keck and Gemini lightcurves by eye. This allowed us to assign an unambiguous cycle number to each minimum and remove any secondary minima. We then determined a linear fit to these times assuming a constant period. This allowed us to determine the period of {\ow} to better than 0.1 sec. We were then able to assign a cycle number to the minima in the lightcurve of the SAAO data obtained in 2016 April and the ULTRACAM data taken in 2016 December. A linear fit to these minima gives an ephemeris of:

\begin{equation}
\begin{aligned}
T_{o} ({\rm HMJD}) = 57695.611284\pm0.000166 \\
+ 0.031015829\pm(7.1\times10^{-8})E \noindent  
\end{aligned}
\end{equation}

We are not able to use the OmegaWhite data taken in March 2012 to refine this further because there is an ambiguity in which of the observed minima are the deepest minima. 

The individual spectra which we obtained for \ow\, all have a relatively low SNR ($\lesssim$10) and hence are not well suited for searching for a radial velocity period. We therefore folded the spectra of \ow\, on the ephemeris shown in equation 1 into 14 phase-bins and co-added individual spectra observed at the same binary phase. This increased the SNR per phase-bin to $\approx$25.

 We used the \texttt{FITSB2} routine \citep{nap04a} to measure the velocities. \texttt{FITSB2} fits Gaussians, Lorentzians and polynomials to the hydrogen and helium lines to fit the continuum, line and line core of the individual lines. Using a $\chi^2$-minimization we fitted the wavelength shifts compared to the rest wavelengths of all suitable spectral lines \citep{gei11a}. Assuming circular orbits, a sine curve was fitted to the folded radial velocity (RV) data points. We find a semi-amplitude $K=422.5\pm21.5$\,\kms\, and a systemic velocity of $\gamma=-14.0\pm11.5\,$\kms\, (Fig.\,\ref{fig:rv_curve}) making it the most compact hot subdwarf binary with the largest radial velocity amplitude. 

\section{Atmospheric parameters of the hot subdwarf star}\label{sec:atmo}

The atmospheric parameters of effective temperature \teff, surface gravity, \logg and helium abundance, $\log{y}$, where $y=n({\rm He})/n({\rm H})$, were determined for the sdOB by fitting the rest-wavelength corrected average spectra with metal-free NLTE model spectra \citep{str07}.  To obtain the average spectrum we shifted the 14 phase-binned spectra which were used for the radial velocity measurement to the rest-wavelength and calculated the combined average spectrum. The spectrum shows Balmer lines as well as neutral (He\,{\sc i}) and ionized (He\,{\sc ii}) helium lines (see Fig.\,\ref{fig:ow0741_aver}). The ionization equilibrium between the He\,{\sc i} and He\,{\sc ii} is most sensitive to the effective temperature of the sdO whereas the broad He\,{\sc ii} and the hydrogen lines in the blue are most sensitive to $\log{g}$. We find \teff=39\,400$\pm$500\,K, \logg=5.74$\pm$0.09 and $\log{y}$=$-$0.14$\pm$0.07 (Fig.\,\ref{fig:ow0741_spec}). The occurrence of both He\,{\sc i} and He\,{\sc ii} as well as the increased helium abundance $\log{y}>-1$ classifies the hot subdwarf as intermediate He-sdOB (see \citealt{heb16}). The errors were derived using a $\chi^2$-minimization. 


\begin{figure*}
\begin{center}
\includegraphics[width=0.47\textwidth]{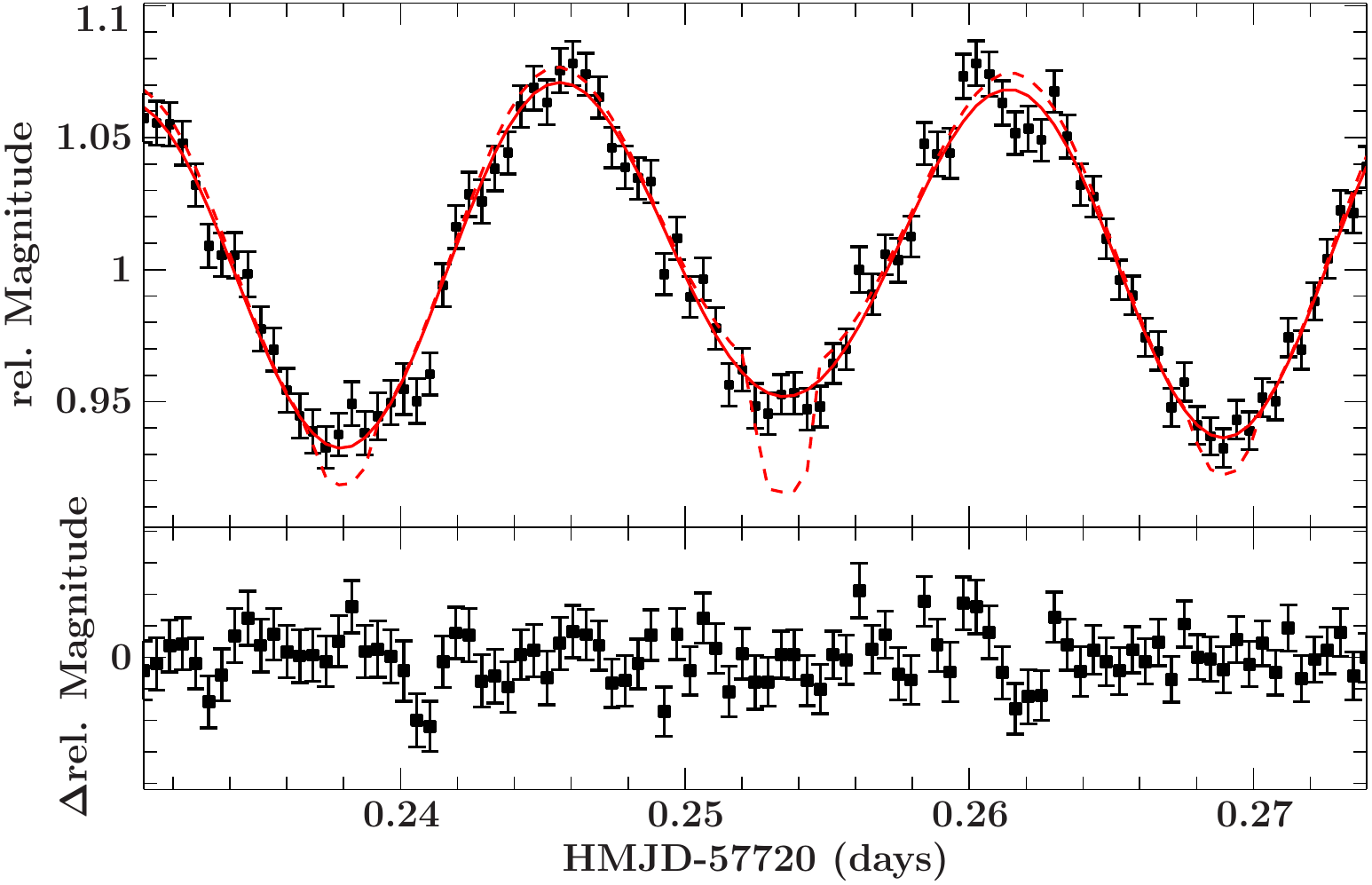}\hspace{0.2cm}
\includegraphics[width=0.47\textwidth]{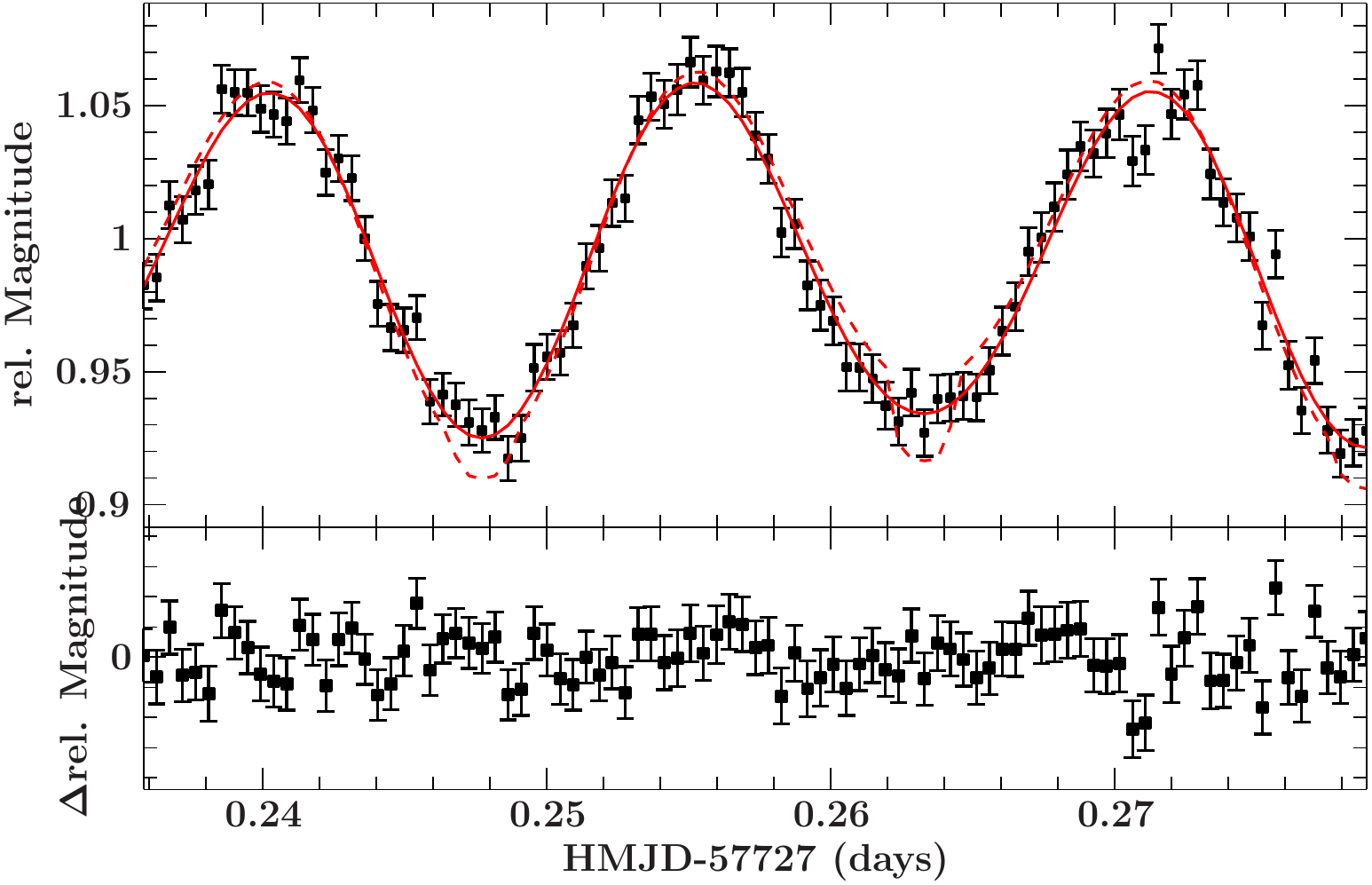}
\includegraphics[width=0.47\textwidth]{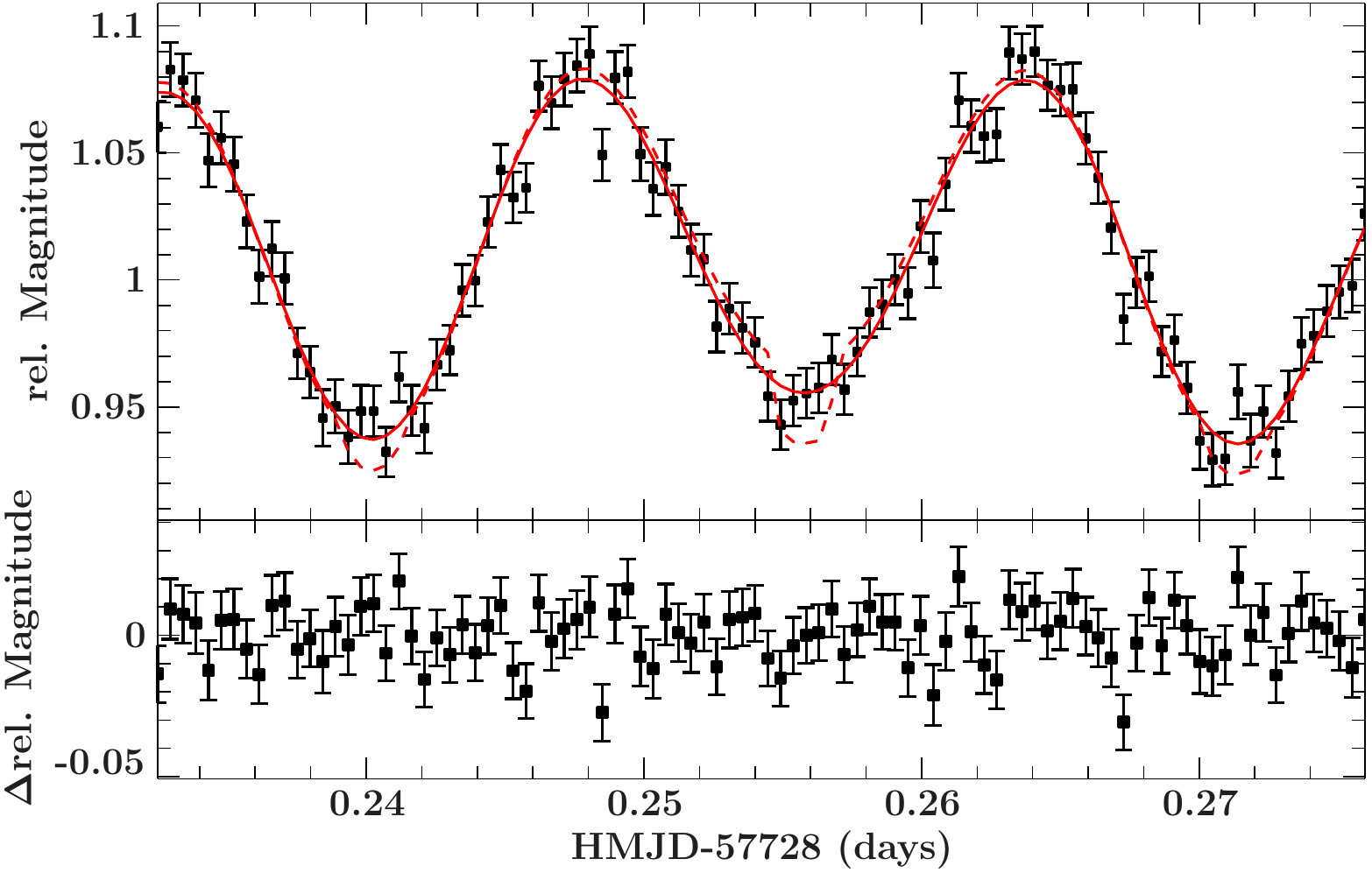}\hspace{0.2cm}
\includegraphics[width=0.47\textwidth]{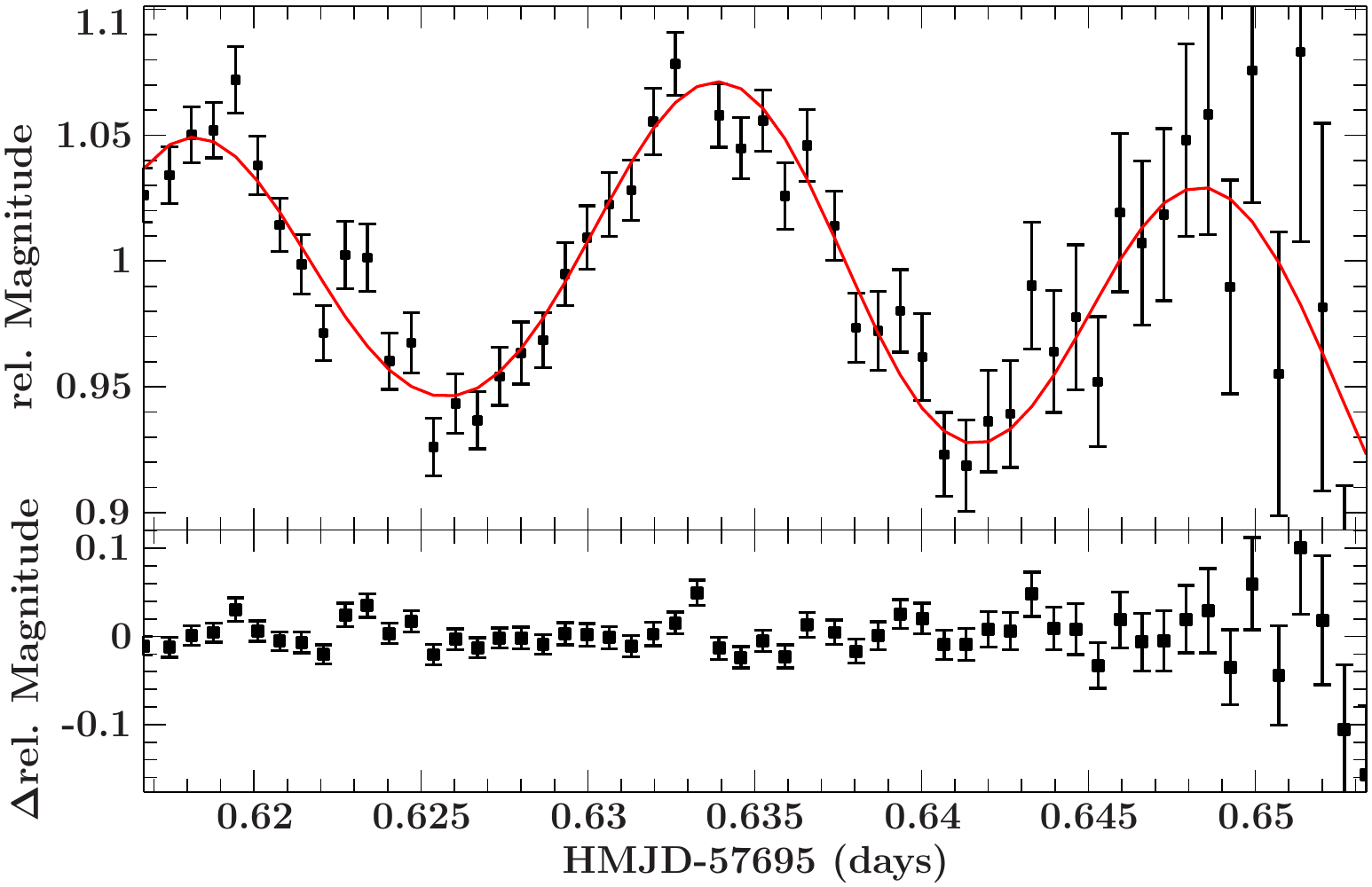}
\includegraphics[width=0.47\textwidth]{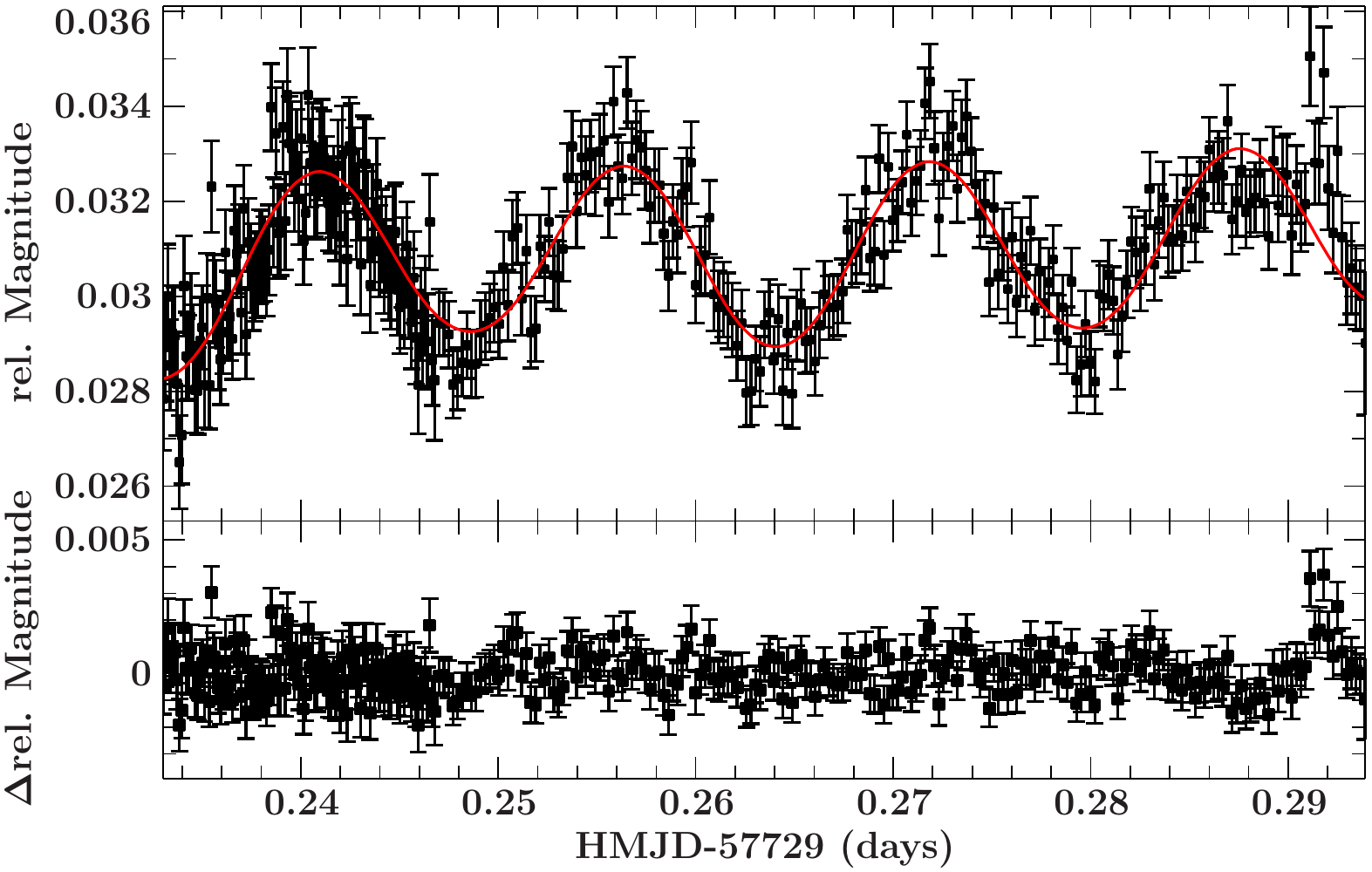}\hspace{0.2cm}
\includegraphics[width=0.47\textwidth]{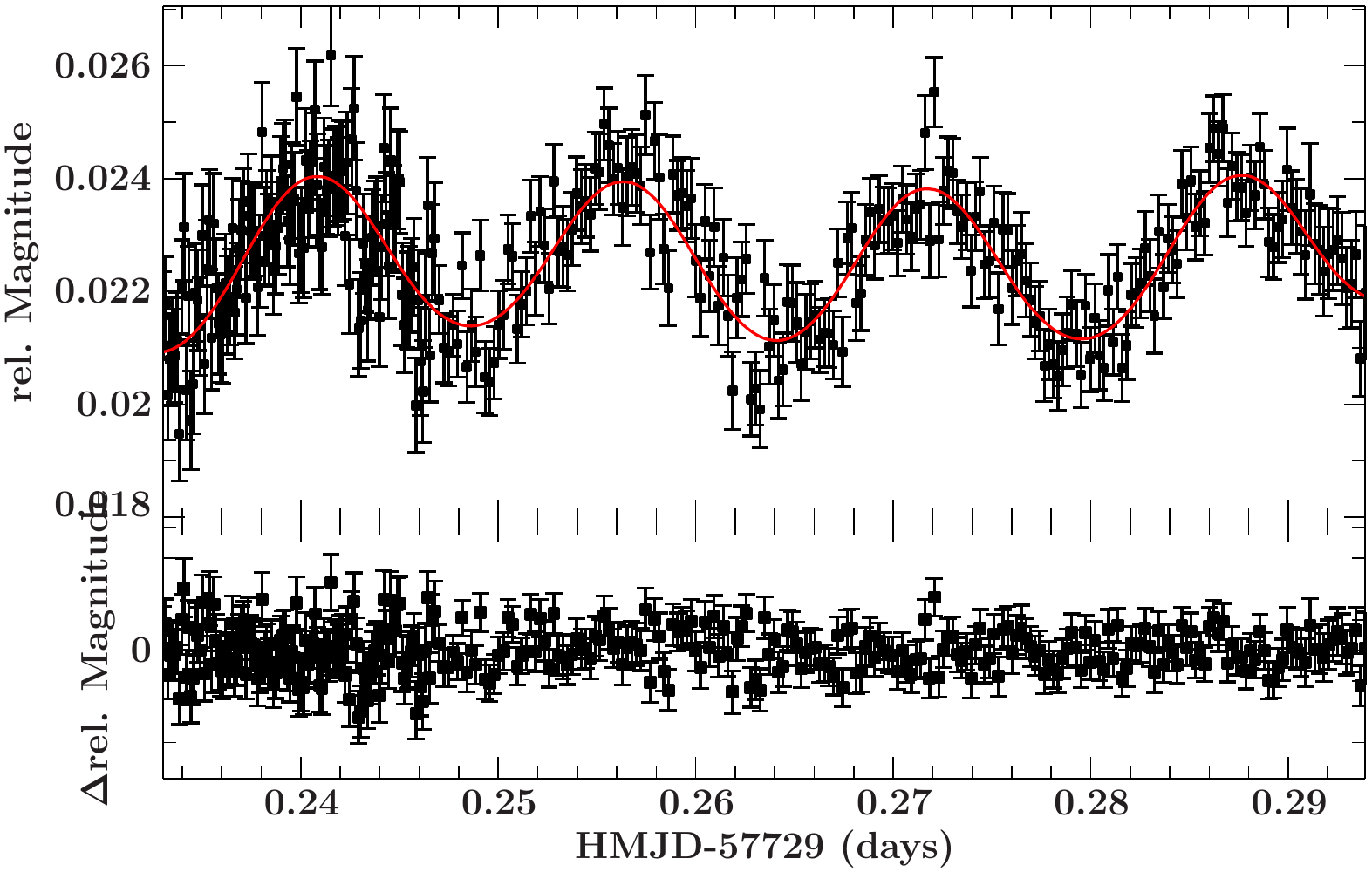}
\end{center}
\caption{Lightcurves shown together with the \texttt{LCURVE} fits. For the Gemini lightcurves the non-eclipsing (solid) and the eclipsing solution (dotted) is shown. The non-eclipsing solution is shown for the Keck and ULTRACAM. The residuals are plotted below. {\bf Upper left:} Gemini (2016-11-28), {\bf Upper right:} Gemini (2016-12-05), {\bf Middle left:} Gemini (2016-12-06), {\bf Middle right:} Keck (2016-11-03), {\bf Lower left:} ULTRACAM \textit{g'} (2016-12-07), {\bf Lower right:} ULTRACAM \textit{r'} (2016-12-07)}
\label{fig:ow0741_model}
\end{figure*} 
 
 \begin{figure*}
\begin{center}
\includegraphics[width=0.33\textwidth]{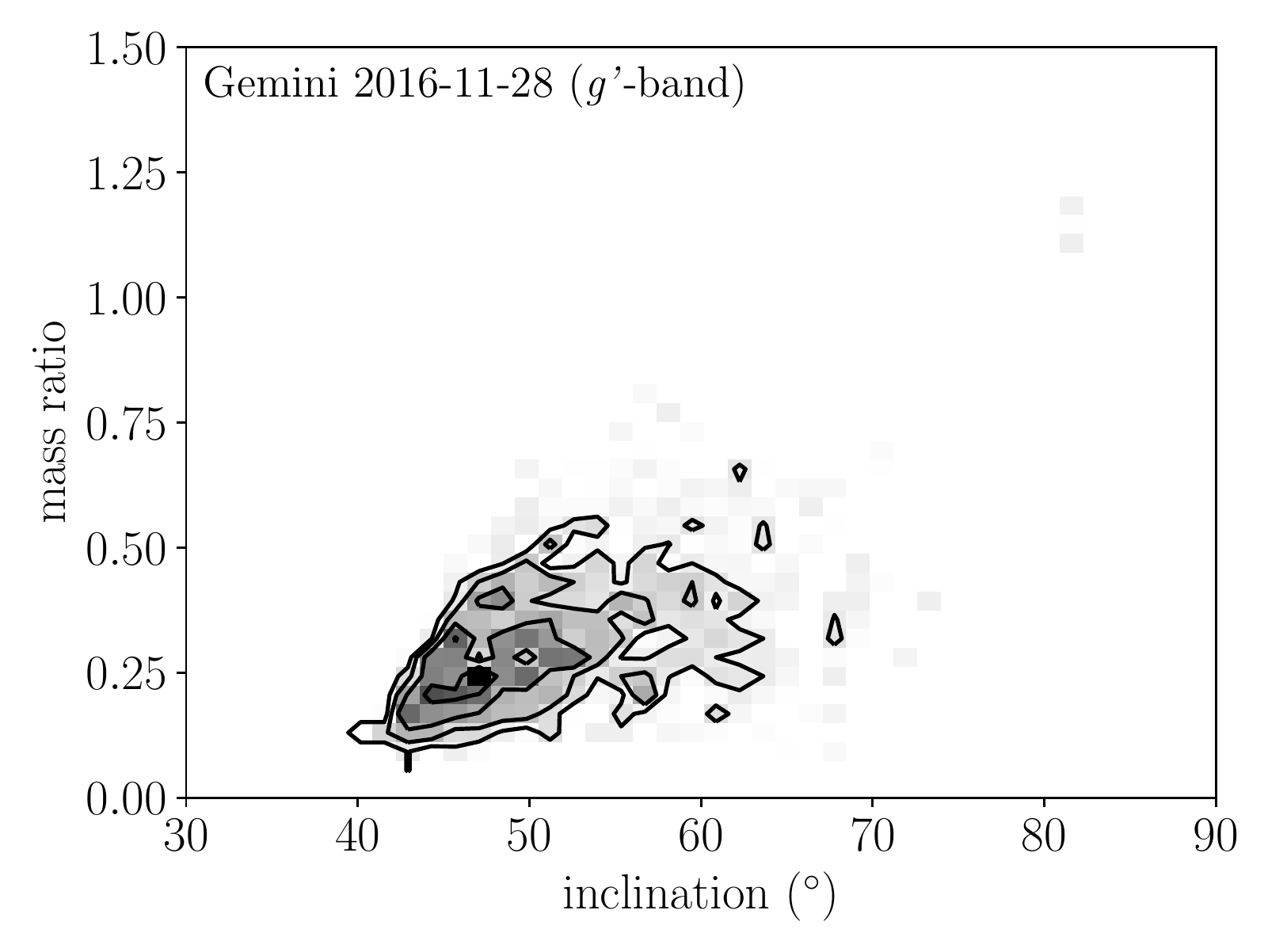}
\includegraphics[width=0.33\textwidth]{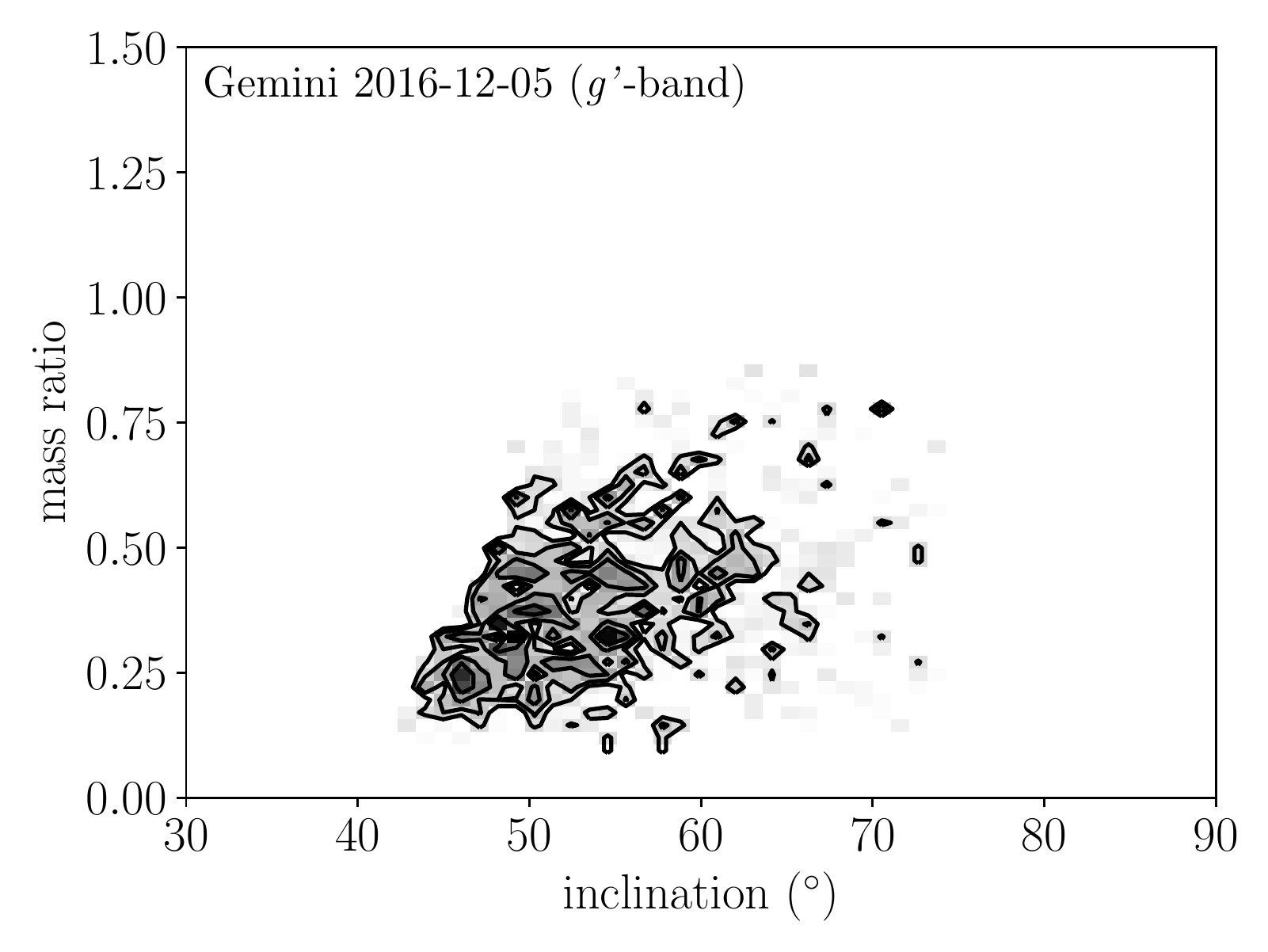}
\includegraphics[width=0.33\textwidth]{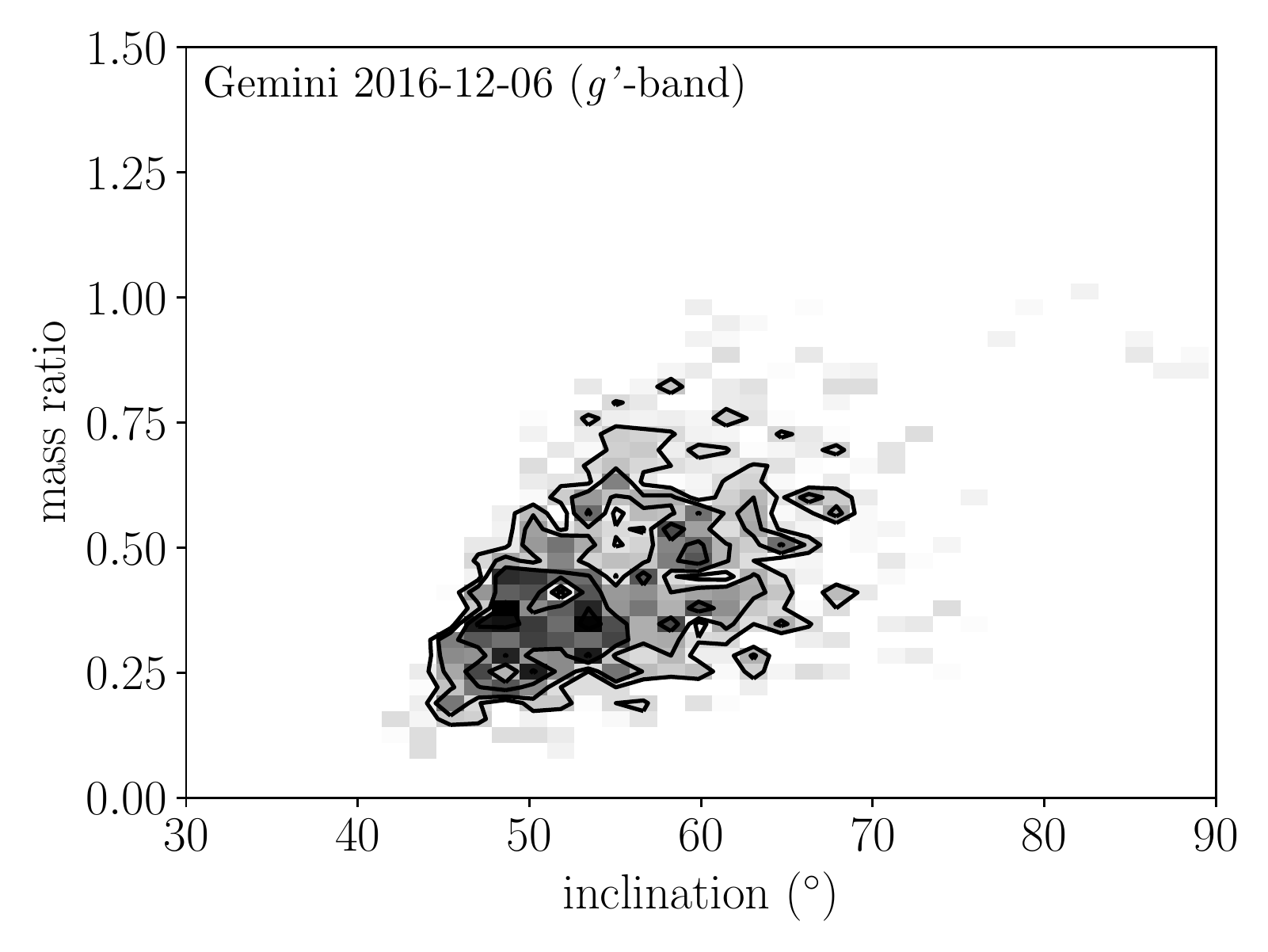}
\includegraphics[width=0.33\textwidth]{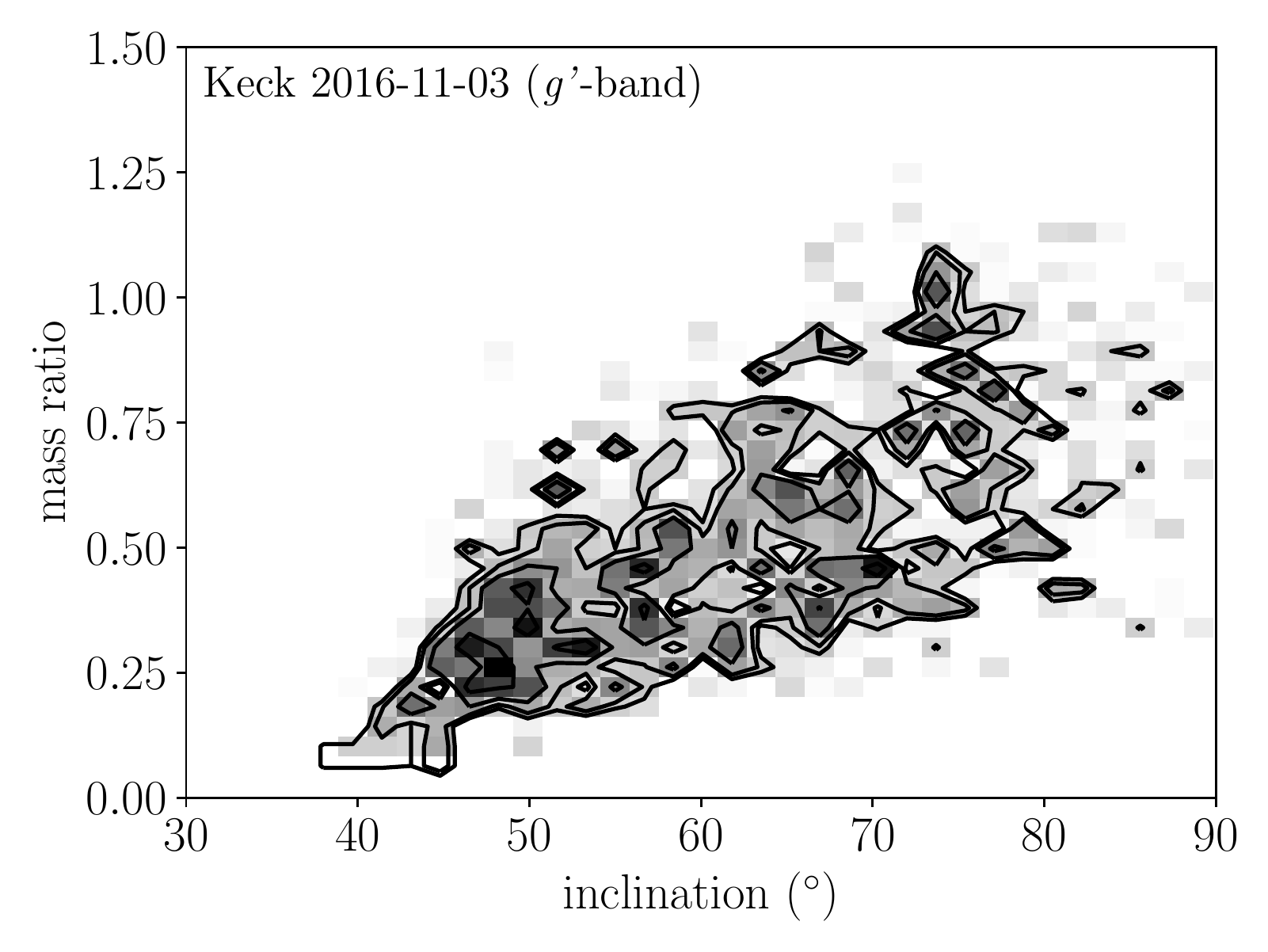}
\includegraphics[width=0.33\textwidth]{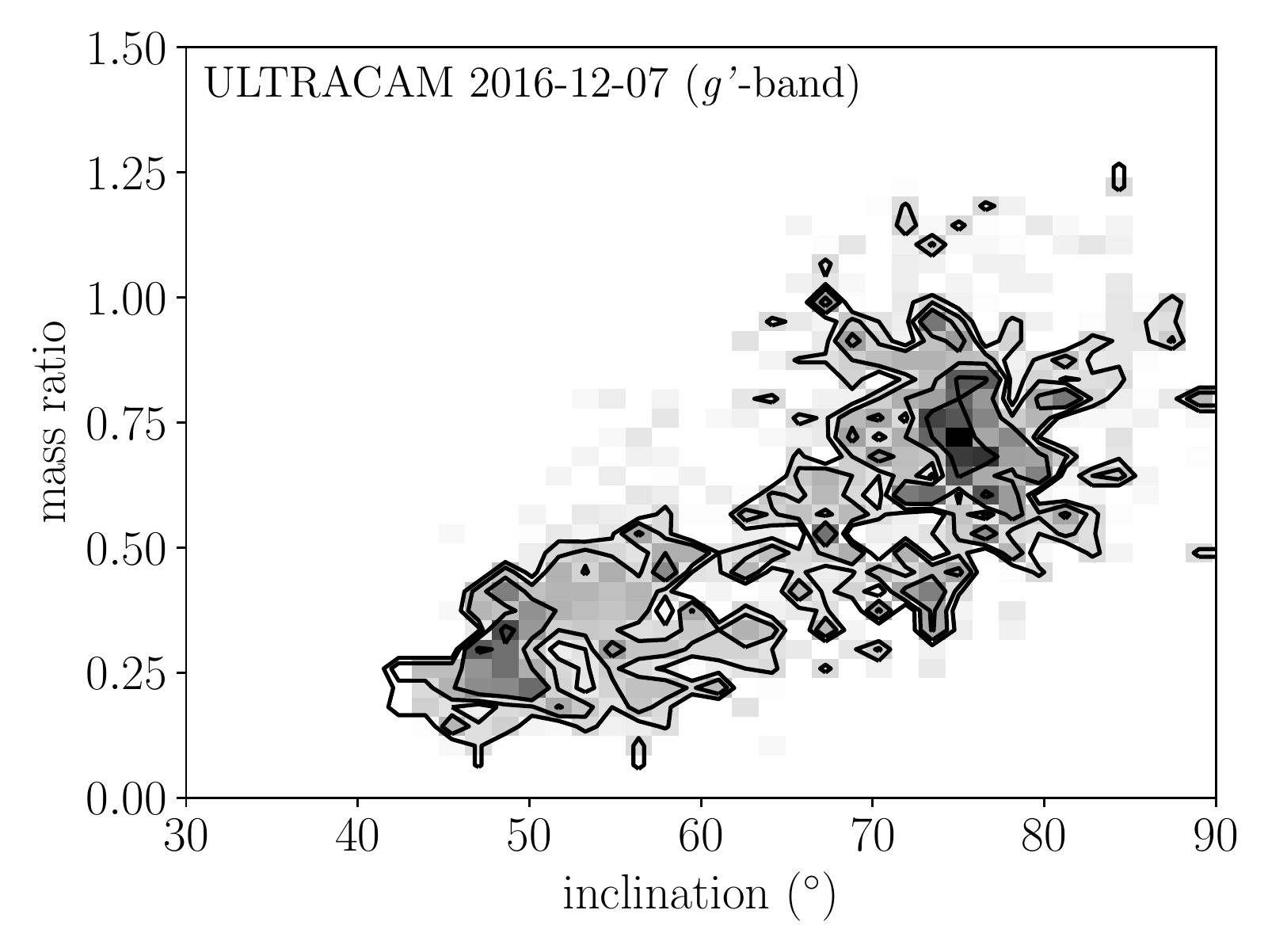}
\includegraphics[width=0.33\textwidth]{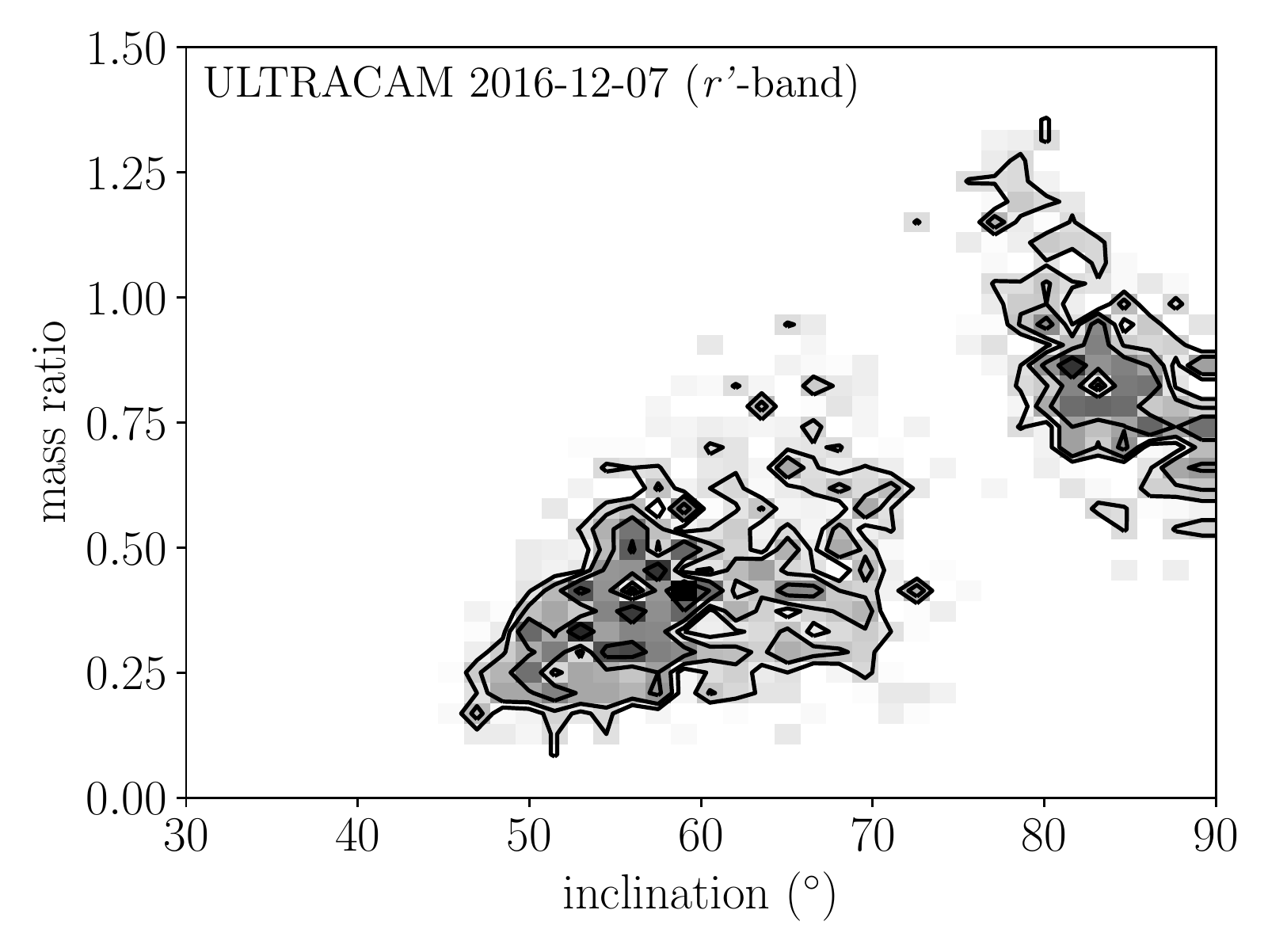}
\end{center}
\caption{2D Histogram of the results from the lightcurve fit. The contours show the 1$\sigma$, 2$\sigma$ and 3$\sigma$ confidence. {\bf Upper left:} Gemini (2016-11-28), {\bf Upper middle:} Gemini (2016-12-05), {\bf Upper right:} Gemini (2016-12-06), {\bf Lower left:} Keck (2016-11-03), {\bf Lower middle:} ULTRACAM \textit{g'} (2016-12-07), {\bf Lower right:} ULTRACAM \textit{r'} (2016-12-07)}
\label{fig:q_i_diag}
\end{figure*}

\section{Lightcurve analysis}
The \texttt{LCURVE} code was used to perform the lightcurve analysis \citep{cop10}. \texttt{LCURVE} uses grids of points to model the two stars. The shape of the stars in the binary is set by a Roche potential. We assume that the orbit is circular and that the rotation periods of the stars are synchronized to the orbital period. The flux that each point on the grid emits is calculated by assuming a blackbody of a certain temperature at the bandpass wavelength, corrected for limb darkening, gravity darkening, Doppler beaming and the reflection effect.

Some additional information was used as input. The orbital period was fixed to $44.66279$\,min as derived in Sec.\,\ref{sec:orbit}. From spectroscopy the \teff, \logg, and the semi-amplitude $K$ was fixed to the values derived in Sec.\,\ref{sec:orbit} and \ref{sec:atmo}. Additionally, as lower limit to the radius (mass) of the white dwarf companion we use the zero-temperature mass-radius relation by Eggleton (quoted from \citealt{ver88}). The passband specific gravity darkening was calculated as described in \citet{blo11}. We use $\mathrm{\beta}=0.33\pm0.01$ in \textit{g'} and $\mathrm{\beta}=0.34\pm0.01$ in \textit{r'}. The limb darkening coefficients were calculated with the Claret limb darkening prescription \citet{cla04}. We used $\mathrm{a_1}=0.613$,  $\mathrm{a_2}=-0.645$, $\mathrm{a_3}=0.621$, and $\mathrm{a_4}=-0.239$ for \textit{g'} and $\mathrm{a_1}=0.422$,  $\mathrm{a_2}=-0.444$, $\mathrm{a_3}=0.475$, and $\mathrm{a_4}=-0.194$ for \textit{r'}. This leaves as free parameters in the model the mass ratio $q = \frac{M_{\rm sdOB}}{M_{\rm WD}}$ , the inclination $i$, secondary temperature $T_{\rm WD}$, the scaled radii of both components $r_{\rm sdOB/WD}$, the velocity scale ($\mathrm[K+K_{\rm WD}]/\sin i$) and the beaming parameter $B$ ($F_\lambda = F_{0,\lambda} \lbrack 1 - B \frac{v_r}{c}\rbrack$, see \citealt{blo11}). In addition, to account for any residual airmass effect we add a third order polynomial.

To determine the uncertainties in the parameters we combine \texttt{LCURVE} with \texttt{emcee} \citep{for13}, an implementation of an MCMC sampler that uses a number of parallel chains to explore the solution space. We use 256 chains and let them run until the chains stabilized to a solution, which took approximately 1000 generations. 

\begin{figure}
\begin{center}
\includegraphics[width=0.48\textwidth]{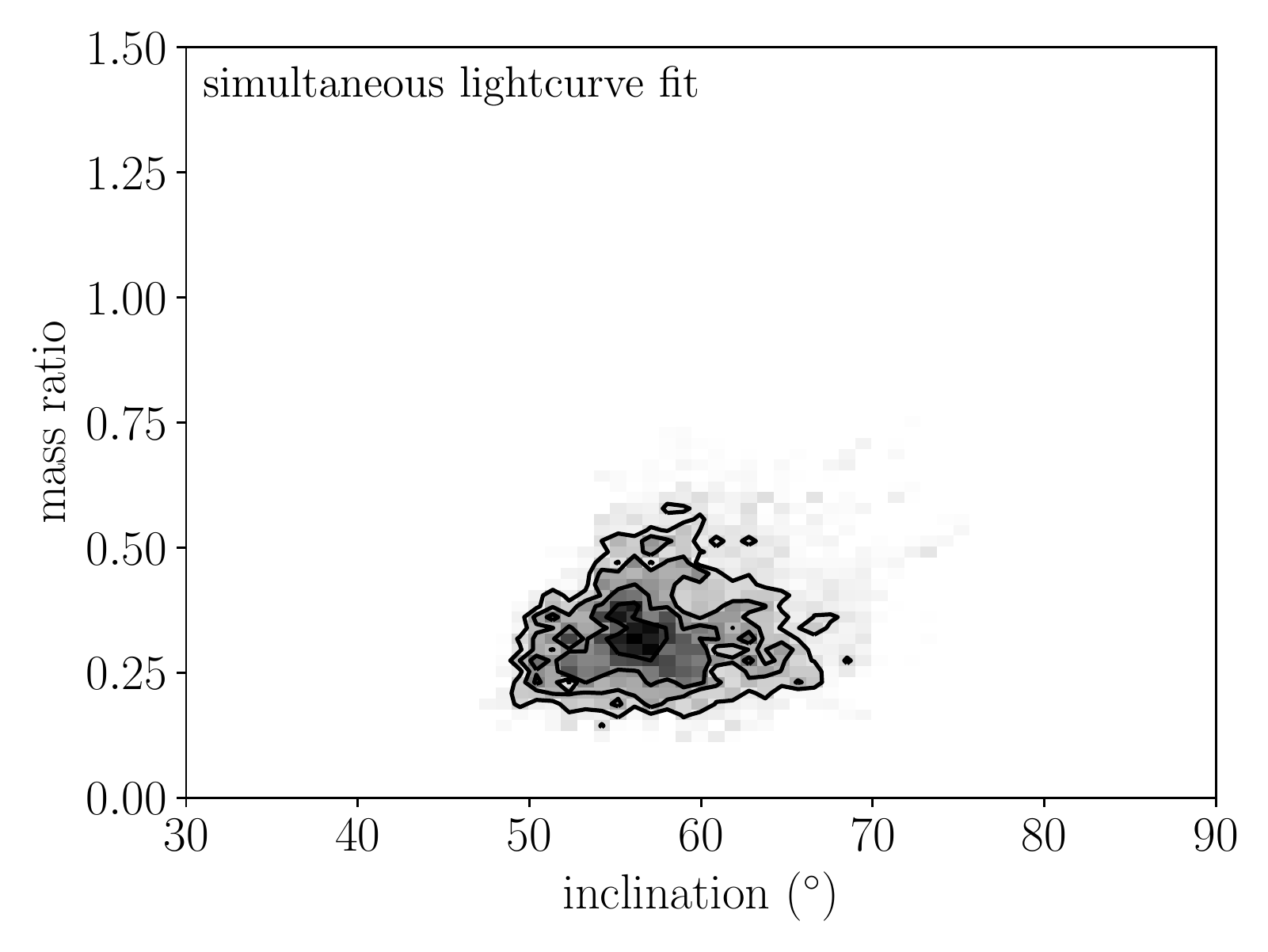}
\end{center}
\caption{2D Histogram of the results from the simultaneous lightcurve fit. The contours show the 1$\sigma$, 2$\sigma$ and 3$\sigma$ confidence.}
\label{fig:q_i_diag_comb}
\end{figure}

For the Keck and ULTRACAM \textit{g'}-band and \textit{r'}-band lightcurves we find two distinct solutions: One which finds the system eclipsing and the other non-eclipsing (Fig.\,\ref{fig:ow0741_model}). The Gemini lightcurves have a higher precision and we find exclusively non-eclipsing solutions (Fig.\,\ref{fig:q_i_diag}). This becomes in particular obvious in the significant larger error bars for inclination ($i$), mass ratio ($q$) and sdOB mass (M$_{\rm sdOB}$) for the individual fits to the Keck and ULTRACM lightcurves (Tab.\,\ref{tab:individual}). To obtain the final solution we fit all six lightcurves simultaneously. From the simultaneous fit we only find non-eclipsing solutions (Fig.\,\ref{fig:q_i_diag_comb}).   
    \begin{table*}
\centering
\caption{Results from the individual lightcurve fits.}
\begin{tabular}{llllllll}
\hline\hline
Lightcurve   & $q = \frac{M_{\rm sdOB}}{M_{\rm WD}}$ & $M_{\rm sdOB}$  & $R_{\rm sdOB}$  & $M_{\rm WD}$  & $i$  & $a$  &  $\frac{R_{\rm sdOB}}{a}$  \\
&   &  [\msol]  & [R$_{\odot}$]  & [\msol] & $[^\circ$] &  [R$_{\odot}$] & \\
  \hline
Keck (2016-11-03) & $0.47\pm0.23$ & $0.40\pm0.25$ & $0.13\pm0.03$ & $0.84\pm0.22$  & $62.3\pm11.5$ & $0.45\pm0.04$  &  $0.30\pm0.04$\\
Gemini (2016-11-28)  & $0.29\pm0.13$ & $0.25\pm0.15$ & $0.12\pm0.02$ & $0.92\pm0.21$  & $50.1\pm6.6$ & $0.44\pm0.04$ &  $0.27\pm0.03$ \\
Gemini (2016-12-05)  & $0.37\pm0.14$ & $0.32\pm0.18$ & $0.13\pm0.02$ & $0.91\pm0.22$  & $53.3\pm6.3$ & $0.45\pm0.04$ & $0.29\pm0.03$ \\
Gemini (2016-12-06)  & $0.39\pm0.16$ & $0.36\pm0.20$ & $0.13\pm0.02$ & $0.92\pm0.21$  & $55.0\pm6.1$ & $0.45\pm0.04$ & $0.29\pm0.03$  \\
ULTRACAM \textit{g'} (2016-12-07) & $0.56\pm0.25$ & $0.44\pm0.26$ & $0.14\pm0.03$ & $0.81\pm0.20$  & $69.6\pm11.4$ & $0.45\pm0.04$  & $0.31\pm0.03$ \\
ULTRACAM \textit{r'} (2016-12-07) & $0.50\pm0.26$ & $0.40\pm0.27$ &  $0.14\pm0.03$ & $0.79\pm0.18$  & $64.4\pm12.8$ & $0.44\pm0.04$  & $0.31\pm0.03$\\
\hline
combined analysis &  $0.32\pm0.10$ & $0.23\pm 0.12$ & $0.11 \pm 0.02$ & $0.72 \pm 0.17$ & $57.4\pm4.7$ & $0.41\pm 0.04$  & $0.28\pm 0.02$ \\
  \hline
\end{tabular}
\label{tab:individual}
\end{table*}

\section{System parameters}
\label{params}

From the strong lightcurve variability caused by ellipsoidal modulation in combination with the radial velocity amplitude and the spectroscopic solution we can solve the system and derive system parameters. Solutions were calculated for each individual lightcurve as well as from a simultaneous fit to all six lightcurves. The results from the latter are taken as the final solution (Tab.\,\ref{tab:individual}).

We find that the system consists of a low mass sdOB with a high-mass WD companion. A mass ratio $q = M_{\rm sdOB}/M_{\rm WD}=0.32\pm0.10$, a mass for the sdOB  $M_{\rm sdOB}=0.23\pm 0.12$\,\msol\,and a WD companion mass $M_{\rm WD}=0.72 \pm 0.17$\,\msol\, were derived. The inclination is found to be $i = 57.4\pm4.7$\,\degree\,(Tab.\,\ref{tab:system}).

We calculate the distance to \ow\, using the visual V magnitude ($m_{\rm V}$), the sdOB mass ($M_{\rm sdOB}$), \teff and \logg as described in \citet{ram01}. The visual V magnitude $m_{\rm V} = 20.06$\,mag was calculated following the conversion from SDSS colors as described in \citet{jes05}. Because \ow\,is located towards the Galactic Bulge significant reddening can occur. \citet{gre15} calculated 3D extinction maps from PanSTARRS data\footnote{http://argonaut.skymaps.info/}  and finds towards the direction of \ow\, an extinction of $E(B-V)=0.34^{+0.032}_{-0.034}$ at distances above $6$\,kpc resulting in a total extinction in the V-band of $A_V = 1.05$\,mag using $R_V = 3.1$. With the corrected magnitude we find a distance of $d=6.6\pm2.1$\,kpc, which for a Galactic latitude of $-3.5^{\circ}$ gives a height below the Galactic plane of $\sim400$\,pc.

For the X-ray observations no source was detected with a 3$\sigma$ upper limit of 0.0036 ct s$^{-1}$. Assuming a 1\,keV thermal bremsstrahlung and a Galactic absorption of 4.4$\times10^{21}$ cm$^{-2}$ (the average to the edge of the Galaxy) we find an upper limit of 2.6$\times10^{-13}$ erg cm$^{-2}$ s$^{-1}$ to the unabsorbed flux (this limit is not strongly dependent on the assumed model). Given the large distance to {\ow}, the upper limit for the X-ray luminosity ($<2 \times10^{33}$ erg s$^{-1}$) is not very constraining since the upper end of X-ray luminosity of hot subwarfs is $\sim10^{31} $ erg s$^{-1}$ \citep{mer16}. In contrast, OWJ0741 was detected in the UVW2 filter (peak effective area 1928\AA) with a corrected count rate of 0.34$\pm$0.01 ct s$^{-1}$ implying a flux 2.04$\pm0.08\times10^{-16}$ erg cm$^{-2}$ s$^{-1}$ \AA$^{-1}$. 

\section{Discussion}

\subsection{Origin of the sdOB star}
Using the constraints from the observations outlined above, we use the stellar evolution code \texttt{MESA} \citep{pax11,pax13,pax15} to put constraints on the nature of the sdOB star.
We then take the most likely model and evolve it forward in time to generate predictions about the future of this system.


\subsubsection{He-star model}
We first test a model with the canonical sdOB mass of $0.462 M_\odot$ with a hydrogen envelope mass of $2.5\times10^{-4} M_\odot$.
We find that after core He burning and during shell He burning, this model spends ${\approx}4$ Myr in crossing the observation box (see dot-dashed purple line in Fig.\,\ref{fig:He-wd-model}).
There are a few problems with the system being in this state.
First, given the \logg, orbital period, and mass ratio, this model fills its Roche lobe (RL) at orbital periods $>50$ minutes, meaning that the higher mass (compared to the derived mass from observations) implies a radius larger than the current RL \footnote{Given the length of the He core burning phase, this system would have a wider orbit during He core burning and avoid Roche Lobe overflow at that phase of evolution}.
Second, the derived mass ratio from observations implies that if the He-star had a mass of $0.46 M_\odot$, the compact companion must be super-Chandrasekhar as well as that the derived mass from the observations is inconsistent with the canonical sdOB mass.


We also tested He-star models between $0.325 M_\odot$ and $0.462 M_\odot$ that experienced He core burning, and found that during the He core burning stage, \teff was too low and \logg was too high to match the measurements.
If the models were able to produce a He-star massive enough to experience He shell burning, this stage reached higher \teff values, but with similar \logg values to the He core burning stage.
Therefore, only He shell burning models near $0.462 M_\odot$ were able to match the \teff and \logg measurements.



\subsubsection{He white dwarf model}

\begin{figure}
\begin{center}
\includegraphics[width=0.49\textwidth]{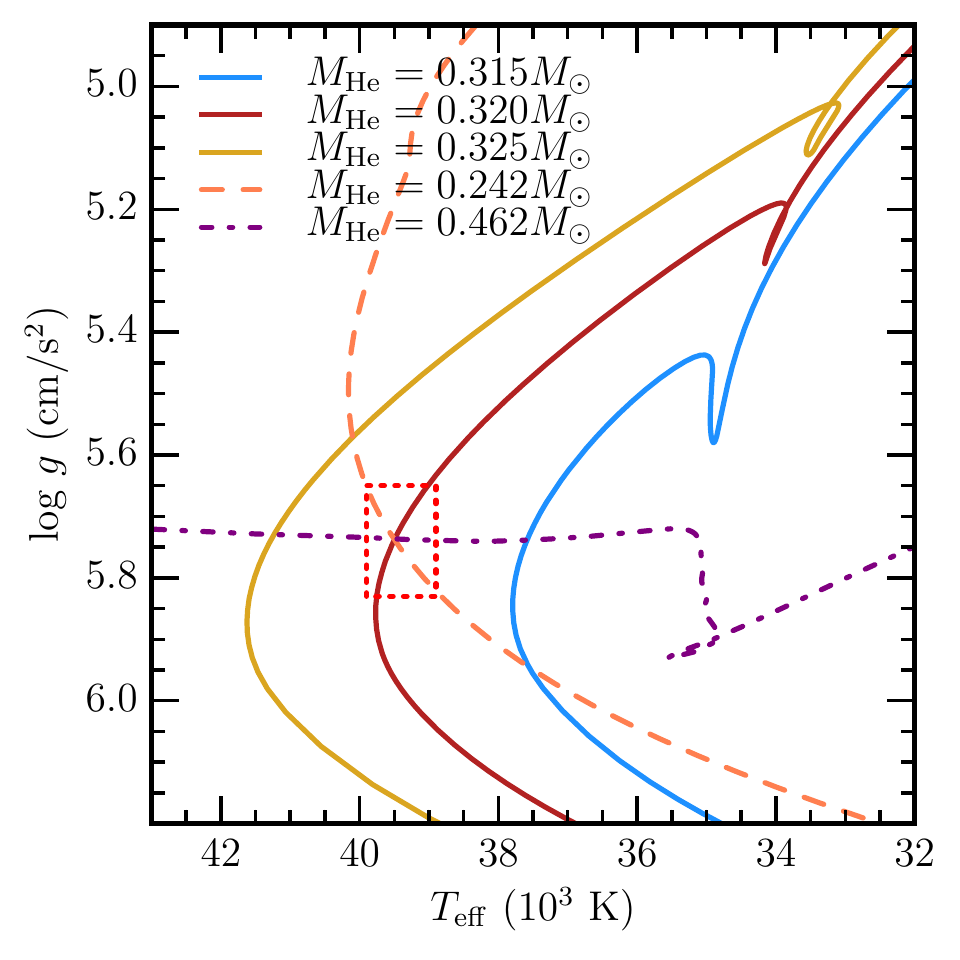}
\end{center}
\caption{The red dotted line marks the observation box from the measurements and error bars given in Table \ref{tab:system}.
The three solid tracks are from He WD models of mass $0.315, 0.320$, and $0.325 M_\odot$ (right to left), where the $0.320 M_\odot$ model spends ${\approx}220,000$ years in the observation box and has a post CE age of ${\approx}1.1$ Myr.
The orange dashed curve is from the $0.242 M_\odot$ model, which starts at the top of the plot, and spends ${\approx}66,000$ years in the observation box and has a post CE age of ${\approx}11$ Myr.
The purple dot-dashed curve is from the $0.462 M_\odot$ model, which starts at the right of this plot and spends ${\approx}4$ Myr in the observation box and has a post CE age of ${\approx}130$ Myr.}
\label{fig:He-wd-model}
\end{figure}

In a second scenario, we consider the sdOB star is a helium white dwarf (He WD) which did not start helium burning. If we assume that the star is a He WD that is just coming out of a CE event, then the \teff and \logg observation box gives a solution of a unique mass \citep{Althaus2013}.
Figure \ref{fig:He-wd-model} shows that this mass must be close to $0.320 M_\odot$, shown by the middle solid curve.
This model spends ${\approx}220,000$ years in the observation box and has a post CE age of ${\approx}1.1$ Myr.

He WDs of this mass range experience diffusion-induced H novae \citep{Althaus2001}, and the tracks of these novae pass through the same \logg  values, but at higher \teff for a given stellar mass.
Therefore, we can construct a lower mass He WD model ($0.242 M_\odot$) that has a too low \teff just out of the CE but passes through the observation box after the first diffusion-induced H-nova.
This is shown by the orange dashed curve in Figure \ref{fig:He-wd-model}.
This model spends ${\approx}66,000$ years in the observation box and has a post CE age of ${\approx}11$ Myr.
Since the $0.320 M_\odot$ He WD model spends more than a factor of 3 longer in the observation box, we conclude that the $0.320 M_\odot$ model is the more likely fit to our observations even though the lower mass He WD model fits the derived mass better.

\begin{table}
\centering
\caption{Overview of the derived parameters for \ow.}
\begin{tabular}{lll}
\hline\hline
Right ascension & RA [hrs]  & 07:41:06.1 \\
Declination  & Dec $[^\circ]$  & --29:48:11.0 \\
Visual magnitude &  $m_{\rm g}$ & 20.02$\pm$0.11  \\
\hline
\multicolumn{3}{l}{\bf{Atmospheric parameters of the sdOB}}  \\ 
Effective temperature & \teff\,[K] & 39\,400$\pm$500 \\
Surface gravity   & \logg  & 5.74$\pm$0.09  \\
Helium abundance& $\log{y}$  & --0.14$\pm$0.06 \\
\hline
\multicolumn{3}{l}{\bf{Orbital parameters}}   \\ 

&  $T_0$ [MHJD]  & $57695.611284\pm1.66\times10^{-4}$  \\
Orbital period & \porb\,[min]  & $44.66279\pm1.16\times10^{-4}$  \\
RV semi-amplitude & $K$ [\kms] & $422.5\pm21.5$ \\
System velocity & $\gamma$\,[\kms] & $-$14.0$\pm$11.5\\\ 
Binary mass function & $f_{\rm m}$ [\msol] &  0.242$\pm$0.020 \\
\hline
\multicolumn{3}{l}{\bf{Derived parameters}} \\

Mass ratio  &  $q = \frac{M_{\rm sdOB}}{M_{\rm WD}}$  & $0.32\pm0.10$  \\
sdOB mass &  $M_{\rm sdOB}$ [\msol] & $0.23\pm 0.12$ \\ 
sdOB radius & $R_{\rm sdOB}$ [R$_{\odot}$] &   $0.11 \pm 0.02$  \\ 
WD mass &  $M_{\rm WD}$ [\msol] & $0.72 \pm 0.17$ \\
Orbital inclination & $i\,[^\circ$] & $57.4\pm4.7$   \\
Separation  & $a$ [R$_{\odot}$]   &  $0.41\pm 0.04$ \\
Distance & $d$ [kpc] & $6.6\pm2.1$ \\
\hline
\end{tabular}
\label{tab:system}
\end{table}

\subsection{Future Predictions}

To predict the future of \ow, we use our best fit model for the sdOB, a $0.320 M_\odot$ He WD, and, given the derived mass ratio, a $0.85 M_\odot$ WD companion.
As mentioned above, WDs in the mass range considered for the sdOB experience diffusion-induced hydrogen novae.
This model experiences its first H-nova 4.1 Myr after passing through the observation box and expands to fill its RL at an orbital period of 40.6 minutes.
Depending on the amount mass loss through Roche Lobe overflow (RLOF) during this hydrogen nova, the model may experience a second hydrogen nova.

The subsequent evolution is governed by gravitational wave radiation which shrinks the orbit. 
Because the WD contraction occurs faster than the orbital decay the He WD begins RLOF after 17.6 Myr, at an orbital period of 5 minutes.
For the given mass ratio, \ow\, is definitely stable according to the dynamical instability \citep{Nelemans2001}, but definitely unstable according to the mass ratio criterion for direct impact accretion. 
Therefore, whether or not this system will be stable to mass transfer depends on the spin-orbit synchronization timescale, $\tau_s$ \citep{mar04} of the accreting WD.
The physics that determine the value of $\tau_s$ is uncertain and previous estimates for systems of this type differ by more than 10 orders of magnitude \citep{Campbell1983,Campbell1984,Fuller2014}.

According to the Figures 1 and 5 from \cite{mar04}, the stability of this system requires a synchronization timescale of $\tau_s \lesssim 10$ yr.
If the synchronization timescale is longer than this, then the WD accretor will extract enough of the orbital angular momentum into its spin angular momentum to cause a merger, leading to an R\,CrB configuration.
If instead the synchronization timescale is short enough, the spin of the accreting WD can couple to the orbit and feed back enough angular momentum to avoid a merger, leading to a stable mass transferring system. 
In this case, the He WD itself will begin transferring degenerate helium at an orbital period of 3 minutes leading to mass transfer rates of ${\approx}3\times10^{-6} M_\odot$/yr.
The accreting WD experiences two small He novae before transitioning to steady He shell burning, leading to the growth of the C/O core.
The mass transfer rate during this phase exceeds the stable burning rate \citep{Brooks2016}, implying mass loss from the binary and an RL filling  accretor.
An intriguing possibility in this phase is the inhibition of direct impact accretion due to the accretor fully filling its RL.
This may enhance the likelihood of stable mass transfer for these systems.
As more mass is transferred, the orbital period increases and the mass transfer rate drops below the minimum steady burning rate and begins mild He flashes that are eventually strong enough to remove mass via RLOF in short ejection episodes.
The last and largest of these flashes \citep{Shen2010} involves only $10^{-2} M_\odot$ and unambiguously remains hydrostatic. 
After the last flash, the accreting WD quiescently grows to $1.04\,M_\odot$, with a $0.94\,M_\odot$ C/O core.

\section{Conclusions and Summary}
\ow\,was discovered as one of bluest variable sources in the OmegaWhite survey. The VST observations revealed variability with a period of 22.6\,min. Follow-up observations show that \ow\, is an ultracompact sdOB binary with a compact companion with \porb=$44.66279\pm1.16\times10^{-4}$\,min making it the most compact hot subdwarf binary known today.

High signal-to-noise ratio photometry obtained with Gemini exclude an eclipse which allows us to put tight constraints on the system parameters. Combining Gemini, Keck and ULTRACAM lightcurves with spectroscopy, we find a mass ratio $q = M_{\rm sdOB}/M_{\rm WD}=0.32\pm0.10$, a mass for the sdOB  $M_{\rm sdOB}=0.23\pm0.12$\,\msol\,and a WD companion mass $M_{\rm WD}=0.72\pm0.17$\,\msol. The derived sdOB mass is inconsistent with the canonical mass for hot subdwarfs of $\approx0.47$\,\msol and therefore the sdOB has not evolved from the standard hot subdwarf channel where the envelope of the subdwarf progenitor gets stripped at the tip of the red-giant branch.

To put constraints on the nature of the sdOB star we compared the derived \teff and \logg to evolutionary tracks for He-stars and He white dwarfs computed with \texttt{MESA}. For the He-star scenario, only a He shell burning star with a mass around $0.462 M_\odot$ is consistent with the derived \teff and \logg. However such a high mass is inconsistent with the derived sdOB mass. For the white dwarf scenario, a He white dwarf with a mass of $0.320$\,\msol\, and a common envelope age of ${\approx}1.1$ Myr is fully consistent with the derived system parameters. Although the evolutionary timescale is about 10 times faster for a He white dwarf, we conclude that the most likely nature of the sdOB is a He white dwarf with a mass of $0.320$\,\msol\, and a common envelope age of ${\approx}1.1$ Myr. 

To study the future evolution of the system we have constructed MESA models, assuming a $0.320$\,\msol\,He white dwarf with a massive white dwarf companion of $0.85$\,\msol\, consistent with the derived mass ratio. 
In 17.6 Myr, the He WD will start RLOF at an orbital period of $\approx$5\,min. Depending on the spin-orbit synchronization timescale the object will either merge to form an R\,CrB star or end up as a stable accreting AM\,CVn type system with a He WD donor. 
Better understanding of the spin-orbit synchronization timescales are required to decide whether the system will merge or prevent the merger. As an AM\,CVn, the system will show weak He-shell flashes, but none strong enough to trigger a double detonation SN\,Ia. Therefore, whether the system merges on contact or not, we conclude that this binary is not a SN Ia progenitor.

\section*{Acknowledgments}
\acknowledgments
This work was supported by the GROWTH project funded by the National Science Foundation under Grant No 1545949. JvR acknowledges support by the Netherlands Research School of Astronomy (NOVA) and the foundation for Fundamental Research on Matter (FOM). 
This research is funded in part by the Gordon and Betty Moore Foundation through Grant GBMF5076 and was also supported by the National Science Foundation under grant PHY 11-25915.
We acknowledge stimulating workshops at Sky House where these ideas germinated. 
 
TRM acknowledge the support from the Science and Technology Facilities Council (STFC),  ST/L00733. SG is supported by the Deutsche Forschungsgemeinschaft (DFG) through grant GE2506/9-1. DK acknowledges financial support from the National Research Foundation of South Africa. PJG acknowledges support from NOVA for the original OmegaWhite observations and hospitality of the Kavli Institute for Theoretical Physics. Armagh Observatory is core funded by the Northern Ireland Executive through the Department for Communities. 

This paper uses observations made at the South African Astronomical Observatory. Based on observations obtained at the European Southern Observatory, proposal 297.D-5010. Some results presented in this paper are based on observations obtained at the Gemini Observatory, proposal ID GS-2016B-FT-9, which is operated by the Association of Universities for Research in Astronomy, Inc., under a cooperative agreement with the NSF on behalf of the Gemini partnership: the National Science Foundation (United States), the National Research Council (Canada), CONICYT (Chile), Ministerio de Ciencia, Tecnolog\'{i}a e Innovaci\'{o}n Productiva (Argentina), and Minist\'{e}rio da Ci\^{e}ncia, Tecnologia e Inova\c{c}\~{a}o (Brazil). The authors thank the staff at the ESO Paranal and Gemini-South observatories for performing the observations in service mode. Some of the data presented herein were obtained at the W.M. Keck Observatory, which is operated as a scientific partnership among the California Institute of Technology, the University of California and the National Aeronautics and Space Administration. The Observatory was made possible by the generous financial support of the W.M. Keck Foundation. The authors wish to recognize and acknowledge the very significant cultural role and reverence that the summit of Mauna Kea has always had within the indigenous Hawaiian community. We are most fortunate to have the opportunity to conduct observations from this mountain. We thank the {\sl Swift} team for the granting our DDT request.

\facilities{Gemini:South (GMOS-S), Keck:I (LRIS), VLT:Antu (FORS2), NTT (ULTRACAM), SALT (SALTICAM), Radcliffe (SHOC)} 
\software{\texttt{LCURVE} \citep{cop10}, \texttt{emcee} \citep{for13}, \texttt{MESA} \citep{pax11,pax13,pax15}, \texttt{FITSB2} \citep{nap04a}}

\bibliographystyle{yahapj}
\bibliography{refs,refs_1508}


\end{document}